\documentclass[a4paper,fleqn,usenatbib]{mnras}

\usepackage[T1]{fontenc}
\usepackage{ae,aecompl}

\usepackage{graphicx}	
\usepackage{amsmath}	
\usepackage{amssymb}	
\usepackage{bm}
\usepackage{color}
\usepackage{ulem}
\usepackage{cancel}

\newcommand{\bq}{\begin{equation}}
\newcommand{\eq}{\end{equation}}
\newcommand{\bqa}{\begin{eqnarray}}
\newcommand{\eqa}{\end{eqnarray}}

\title[Convection and pulsation stability of stars]{Turbulent convection and pulsation stability of stars --
II. Theoretical instability strip for $\delta$ Scuti and $\gamma$ Doradus stars}

\author[D. R. Xiong et al.]{
D. R. Xiong,$^{1}$
L. Deng,$^{2}$
C. Zhang$^{2}$
and K. Wang$^{2}$
\\
$^{1}$Purple Mountain Observatory, Chinese Academy of Sciences, Nanjing 210008, China\\
$^{2}$Key Laboratory of Optical Astronomy, National Astronomical Observatories, Chinese Academy of Sciences, Beijing, 100101, China
}

\date{Accepted 2016 January 11. Received 2016 January 10; in original form 2015 August 4}

\pubyear{2015}

\begin{document}
\label{firstpage}
\pagerange{\pageref{firstpage}--\pageref{lastpage}}
\maketitle

\begin{abstract}
By using a non-local and time-dependent convection theory, we have calculated radial and low-degree non-radial oscillations for stellar evolutionary models with $M=1.4$--3.0\,$\mathrm{M}_\odot$. The results of our study predict theoretical instability strips for $\delta$ Scuti and $\gamma$ Doradus stars, which overlap with each other. The strip of $\gamma$ Doradus is slightly redder in colour than that of $\delta$ Scuti. We have paid great attention to the excitation and stabilization mechanisms for these two types of oscillations, and we conclude that radiative $\kappa$ mechanism plays a major role in the excitation of warm $\delta$ Scuti and $\gamma$ Doradus stars, while the coupling between convection and oscillations is responsible for excitation and stabilization in cool stars. Generally speaking, turbulent pressure is an excitation of oscillations, especially in cool $\delta$ Scuti and $\gamma$ Doradus stars and all cool Cepheid- and Mira-like stars. Turbulent thermal convection, on the other hand, is a damping mechanism against oscillations that actually plays the major role in giving rise to the red edge of the instability strip. Our study shows that oscillations of $\delta$ Scuti and $\gamma$ Doradus stars are both due to the combination of $\kappa$ mechanism and the coupling between convection and oscillations, and they belong to the same class of variables at the low-luminosity part of the Cepheid instability strip. Within the $\delta$ Scuti--$\gamma$ Doradus instability strip, most of the pulsating variables are very likely hybrids that are excited in both p and g modes.
\end{abstract}




\section{Introduction}\label{sec1}

On the Hertzsprung--Russell (H--R) diagram, the diagonal Cepheid instability strip intersects with the main sequence (MS) at A--F dwarfs, along which a few types of classical variables are confined: Cepheids of Population I and II are located in the upper parts, RR Lyrae stars are in the middle part, while $\delta$ Scuti stars and white dwarfs are found near the MS and even lower luminosities. All of them are excited by $\kappa$ mechanism due to the ionization zones of hydrogen and helium. The overall properties for variables in the strip are that giants tend to pulsate in a single or a few low-order radial modes with large amplitudes, while dwarfs usually oscillate in multiple radial and non-radial modes with rather small amplitudes. Only very few of $\delta$ Scuti stars have relatively large amplitude in radial modes. High-amplitude $\delta$ Scuti stars are normally slow rotators ($v\sin i\leq 30\,\mathrm{km\,s}^{-1}$), while other stars in the same region of H-R diagram normally have high rotation speed ($v\sin i \sim 150\,\mathrm{km\,s}^{-1}$). It is believed that rotation is responsible for the existence of two types of $\delta$ Scuti stars. Whether this is true or not is still to be solved.

The study of $\delta$ Scuti stars has a history of over a hundred years since the first discovery by Campbell \& Wright (1900). Research in this field has been largely boosted since the success of helioseismology, and it  has then been the first objective in asteroseismology therefore has made a great leap forward. The earlier work has been intensively reviewed by Breger (2000). In contrast to that, the study of $\gamma$ Doradus stars, only 20 yr after officially classified (Balona, Krisciunas \& Cousins 1994), has just been started. On the H--R diagram, they are a group of variable stars attached to the low-temperature region outside the red edge of the $\delta$ Scuti strip, being the first class of g-mode oscillators in the Cepheid instability strip. The so-called convective blocking (Guzik et al. 2000) is believed to be the excitation mechanism of oscillations, with the original idea coming from the hypothesis of `frozen-in convection' (Warner, Kaye \& Guzik 2003). However, such a hypothesis can never give rise to the red edge of the instability strip theoretically. Using a time-dependent mixing-length theory, it was confirmed that convective flux blocking could be the driving mechanism in $\gamma$ Doradus stars, given the mixing-length parameter was fine-tuned (Dupret et al. 2005).

Convection drives transportation of energy and momentum in stellar interiors, and therefore affects the structure and oscillation stabilities. For low-temperature stars with extended convective envelope, convection overtakes radiation and becomes the major play of energy transportation. The dynamic and thermodynamic couplings between convection and oscillations become the major mechanisms of excitation and damping against oscillations. Based on hydrodynamic equations and the theory of turbulence, we developed a non-local and time-dependent theory of convection (Xiong 1989; Xiong, Cheng \& Deng 1997; Xiong, Deng \& Zhang 2015). The current paper is the second of the series of work on turbulent convection and stability of pulsating stars. In this work, we calculated linear non-adiabatic oscillations in radial and low-degree ($l=1$--4) non-radial modes for evolutionary models of $1.4$--3.0\,$\mathrm{M}_\odot$ stars, from which the blue and red edges of $\delta$ Scuti and $\gamma$ Doradus stars were defined. Such calculations also facilitate studies of the properties of the strips and the relation between them. The {\it Kepler} mission provided high-precision photometry in continuous long time baseline for over 100 000 stars, among which about 1500 new $\delta$ Scuti and $\gamma$ Doradus stars were discovered. This high-precision sample not only greatly enhanced the data base, but also offers a best opportunity for the studies of the two types of variables (Balona \& Dziembowski 2011; Balona et al. 2011). In Section~\ref{sec2}, we discuss the general properties of linear non-adiabatic oscillations in our theoretical scheme. The theoretical instability strips of $\delta$ Scuti and $\gamma$ Doradus stars defined based on our calculations are presented in Section~\ref{s3}. Thorough statistics of the observational properties of these variables and comparison with existing theoretical results are given in Section~\ref{s4}. The excitation and stabilization mechanisms and related problems are discussed in Section~\ref{s5}. The work is then summarized together with some discussions in the last section.

\section{Calculations of linear non-adiabatic oscillations for evolutionary models of stars}\label{sec2}

In the first paper of this series (Xiong et al. 2015, P1 hereafter), starting from hydrodynamic equations and using the classical Reynolds method for turbulence and other achievements in this field, we derived a set of average hydrodynamic equations and a set of dynamic equations of auto- and cross-correlation functions of turbulent velocity and temperature. Combination of the two sets forms complete and closed dynamic equations that are ready for calculations of stellar structure and oscillations (equations (33)--(41) in P1).  Applying the equations, we calculated linear non-adiabatic oscillations in radial (p0--p39) and low-degree ($l$=1--4) non-radial g29--p29 modes of evolutionary stellar models from zero-age main sequence (ZAMS) to red giant branch (RGB) for stars with $M$=1.4--3.0\,$\mathrm{M}_\odot$. All the numerical work is divided into three categories in the following.

\begin{enumerate}
\item Stellar evolutionary models: evolutionary tracks of stars with $M=1.4$--3.0\,$\mathrm{M}_\odot$, abundance $X=0.70$, $Z=0.02$ with solar metal mixture (Grevesse \& Noels 1993), and an overshooting parameter $l_\mathrm{ov}=0.5H_P$ were followed from ZAMS to RGB using Padova code (Bressan et al. 1993, with updated input physics).
\item Calculations of equilibrium models of non-local convective stellar envelopes.

The primary excitation and damping come from the outer convective stellar envelopes, those coming from the core regions hardly contribute. The main goal of the current work is to study the stability of stellar oscillations, therefore, it is efficient and equivalent to calculate only the convective envelopes compared with full-scale stellar models. Setting all the terms containing velocity $u^i$ and its time derivatives in the dynamic equations for calculations of stellar structure and oscillations to zeros, we can have a set of 12 equations to calculate stellar envelope structure in non-local convection.

Under local convection theory, the envelope structure of stars can be dealt with by a set of four normal differential equations as an initial value problem.  Given the temperature $T$ , pressure $P$, radius $r$, and luminosity $L_r$ at the surface $M_r=M_0$, one can integrate inwards to a pre-set bottom boundary at $M_r=M_\mathrm{b}$. In dealing with the envelope structure under non-local convection, however, convection quantities all have boundary conditions at both the surface and bottom, therefore, it is a boundary problem of a set of 12 normal differential equations. The upper boundary is set at a layer with shallow enough optical depth, say $\tau_0=10^{-3}$ for instance, in our calculations. While at the surface, ($r$, $T$, $P$, $L_r$)$_0$ are given by iteration method, which make the following equation hold for the photosphere ($\tau=2/3$):
\bq
L=4\mathrm{\pi} r^2_\mathrm{ph}\sigma T_\mathrm{e}^4,
\label{eq1}
\eq
where $T_\mathrm{e}$ is the effective temperature, $r_\mathrm{ph}$ is the radius ($\tau=2/3$) of the photosphere. The lower boundary of the envelope models, on the other hand, is set at a deep enough layer where the temperature is about $8 \times 10^6$\,K or fractional radius $r_\mathrm{b}/R_0 \sim 0.03$--$0.05$, so that the envelope model has its both boundaries sitting in the overshooting regions and all the convective unstable zones are included. The boundary conditions for the convection quantities are then derived using the asymptotic analysis of convective overshooting (Unno, Kondo \& Xiong 1985; Deng, Xiong \& Chan 2006).

The boundary problem of the current set of normal differential equations adopted the Henyey's algorithm (Henyey, Forbes \& Gould 1964), which needs initial testing solutions. The initial testing solutions are provided in the following way: starting from a local convection theory, an envelope model is constructed in which convection quantities vanish in the local convective stable zones; the solutions are then expanded to overshooting zones using the properties of the analytical asymptotic solution of non-local convective overshooting, such that a quasi-non-local convection model is built and then applied as the initial solutions in the Henyey iterative process.

There is a singularity in computational mathematics sense emerging in the equations of convective envelope under non-local convection theory, which makes the initial testing solutions very delicate, or say the convergence domain is made rather small. As a result, special care is needed in getting the right numerical solutions. Just as we demonstrated in P1, using the quasi-anisotropic convection model (QACM) to replace the completely anisotropic convection model (ACM) is in fact very practical. In terms of both thermal ($T$--$P$) structure and the fields of turbulent velocity and temperature, QACM and ACM are very close (see Figs. 4 and 5 in P1). Non-adiabatic stellar oscillations are calculated using both ACM and QACM as equilibrium solutions, the results delivered in the two schemes are nearly identical (see Figs. 6 and 7 in P1). Although there are only 2 equations less in QACM compared to that in ACM, the former behaves far better in terms of stability and convergence, therefore is highly recommended in practice.

\item Calculations of the linear non-adiabatic oscillations of equilibrium models of stars (ACM/QACM)

Linearizing the dynamic equations for calculations of stellar structure and oscillations (equations (33)--(41) in P1), we have the equations for linear non-adiabatic oscillation calculations. For radial oscillations of stars, this is a set of 12 linear differential equations of complex quantities, while for non-radial oscillations it becomes 16. Henyey's method is also applied in the integrations of the linear non-adiabatic oscillations. Boundary conditions at the surface and bottom are given by Unno et al. (1989), and convective boundary conditions are based on the asymptotic solutions of convective overshooting (Xiong, Cheng \& Deng 1998). In our non-local and time-dependent convection theory, the troublesome spatial oscillations of thermal variables in the calculation of non-adiabatic oscillations using local time-dependent convection theories (Keeley 1977; Baker \& Gough 1979; Gonczi \& Osaki 1980) are largely suppressed or even completely eliminated.

\end{enumerate}

\section{Instability strips for $\delta$ Scuti and $\gamma$ Doradus stars}\label{s3}

Using the numerical scheme described in previous section, we calculated radial and low-degree ($l=1$--4) non-radial oscillations of stars with $M=1.4$--3.0\,$\mathrm{M}_\odot$ from ZAMS up to red-giant phase. In the context of the current work, we present results only for models before entering the giant phase, i.e. $\log T_\mathrm{e}\geq 3.7$. The results of RGB phase are related to oscillations of different types of stars and will be presented in a further work which is still in progress.

A slightly modified version of MHD equation of state (Hummer \& Mihalas 1988; Mihalas, D\"appen \& Hummer 1988; D\"appen et al. 1988), and OPAL tabular opacity (Rogers \& Iglesias 1992) complemented by low-temperature tabular opacity (Alexander \& Ferguson 1994) have been adopted.

\subsection{The $\delta$ Scuti instability strip}\label{s31}

\begin{figure}
\includegraphics[width=\columnwidth]{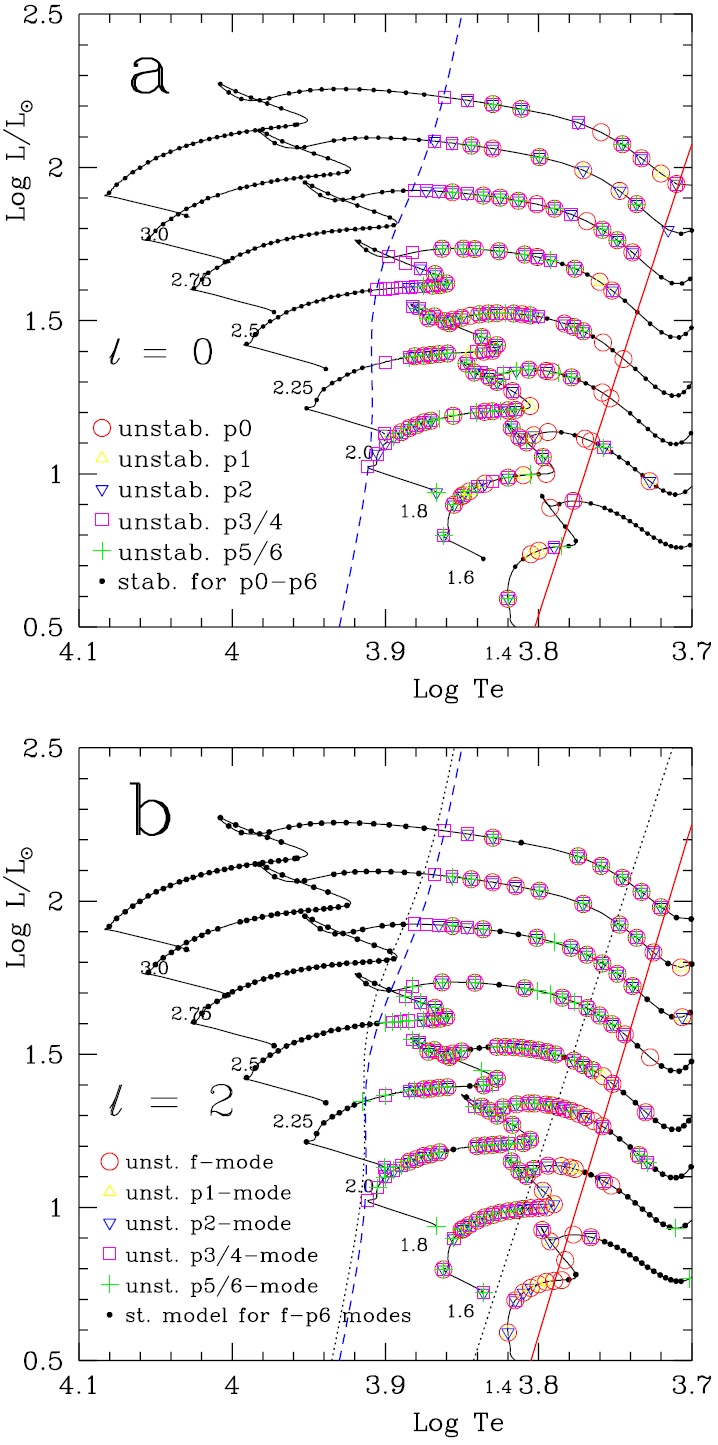}
\caption{Pulsationally stable and unstable low-order p modes on the H-R diagram. The small dots represent pulsationally stable modes; the open circles, triangles, inverse triangles, squares, and plus signs are the unstable modes. The dashed and solid lines are respectively the theoretical blue and red edges of the instability strip for low-order modes. Panel a: radial modes. Panel b: non-radial modes with $l=2$. The dotted lines in panel b are the theoretical blue and red edges of the radial modes as in panel a.}
\label{fig1}
\end{figure}

Fig.~\ref{fig1} shows the distributions of the theoretical radial ($l=0$; panel a) and non-radial ($l=2$; panel b) stable (small dots) and unstable (other symbols) p modes on the H-R diagram. The dashed and solid lines are the blue and red edges of the $\delta$ Scuti instability strip from our calculations, which can be described by two lines approximately as the following.

For low-order radial modes (p0--p6),

\begin{subequations}
\begin{align}
{\rm blue~edge:~} \log T_\mathrm{b}&=3.910-0.040\left(\log\frac{L}{\mathrm{L}_\odot}-1.0\right)\nonumber\\
&+0.02\exp\left[-7\left(\log\frac{L}{\mathrm{L}_\odot}-1.6\right)^2\right],\\
{\rm red~edge:~}  \log T_\mathrm{r}&=3.770-0.065\left(\log\frac{L}{\mathrm{L}_\odot}-1.0\right);
\end{align}
\label{eq2}
\end{subequations}
\\and for low-order non-radial modes (f--p6),

\begin{subequations}
\begin{align}
{\rm blue~edge:~} \log T_\mathrm{b}&=3.910-0.040\left(\log\frac{L}{\mathrm{L}_\odot}-1.0\right)\nonumber\\
&+0.02\exp\left[-7\left(\log\frac{L}{\mathrm{L}_\odot}-1.5\right)^2\right],\\
{\rm red~edge:~}  \log T_\mathrm{r}&=3.775-0.060\left(\log\frac{L}{\mathrm{L}_\odot}-1.0\right).
\end{align}
\label{eq3}
\end{subequations}

\subsection{The $\gamma$ Doradus instability strip}\label{s32}
\begin{figure}
\includegraphics[width=\columnwidth]{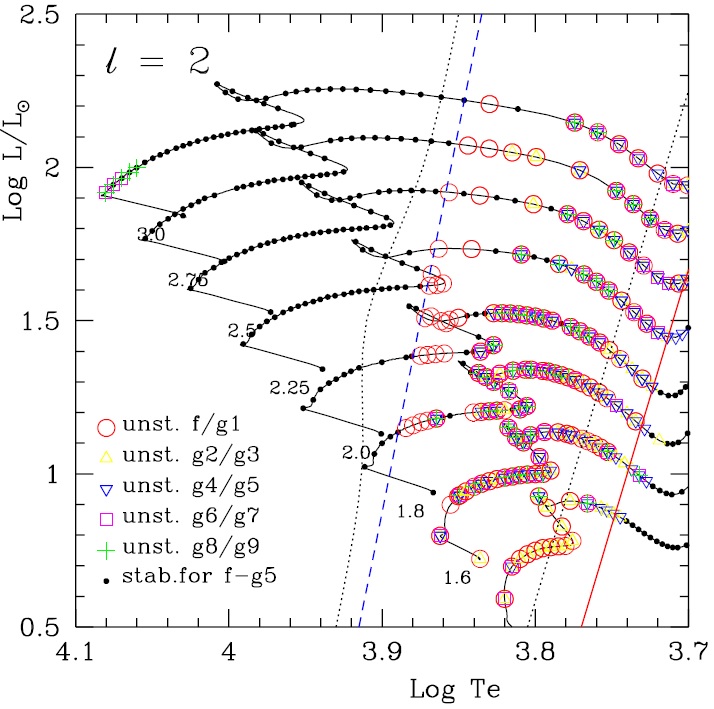}
\caption{Pulsationally stable and unstable g modes on the H-R diagram for $l=2$. The small dots are the stable f--g5 modes. The open circles, triangles, inverse triangles, squares, and plus signs are the unstable f--g9 modes. The dashed and solid lines are respectively the theoretical blue and red edges of the instability strip for the f--g9 modes. The dotted lines are the theoretical blue and red edges for low-order non-radial p modes.}
\label{fig2}
\end{figure}

Fig.~\ref{fig2} depicts the distributions of pulsationally stable (small dots for g1--g5) and unstable (other symbols) low-order g modes with $l=2$ on the H-R diagram. The dashed and solid lines are the corresponding blue and red edges of the instability strip for f--g9 modes that can be represented approximately by two straight lines (independent of $l$):
\begin{subequations}
\begin{align}
{\rm blue~edge:~} \log T_\mathrm{b}&=3.895-0.040\left(\log\frac{L}{\mathrm{L}_\odot}-1.0\right),\\
{\rm red~edge:~}  \log T_\mathrm{r}&=3.740-0.060\left(\log\frac{L}{\mathrm{L}_\odot}-1.0\right).
\end{align}
\label{eq4}
\end{subequations}

The dotted lines are the blue and red edges of the instability strip for low-order non-radial p modes as in Fig.~\ref{fig1}b.  Just as indicated by observations that $\delta$ Scuti and $\gamma$ Doradus instability strips actually overlap partially on the colour-magnitude diagram, with the later being slightly redder systematically. {\it Kepler} observations reveal the fact that p and g modes are actually co-existing in these variable stars,  therefore they are $\delta$ Scuti-$\gamma$ Doradus twins. Such a fact gives a clearance for the conjecture that stars in this instability region are cousins or twins with both $\delta$ Scuti and $\gamma$ Doradus properties suggested by ground-based observations for a long time.

\subsection{Dependence of pulsation instability on $l$}\label{s33}
\begin{figure}
\includegraphics[width=\columnwidth]{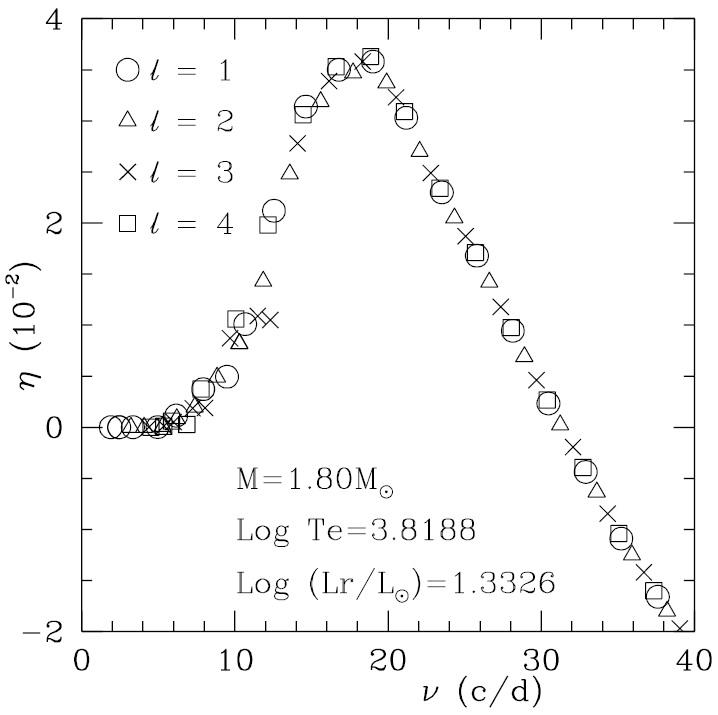}
\caption{Pulsation amplitude growth rate $\eta$ versus the oscillation frequency $\nu$ for a stellar model ($M=1.8\,\mathrm{M}_\odot$, $\log L/\mathrm{L}_\odot=1.3326$, $\log T_\mathrm{e}=3.8188$). The circles, triangles, crosses, and squares are respectively the non-radial oscillation modes with $l=1$--4.}
\label{fig3}
\end{figure}

Fig.~\ref{fig3} presents the pulsation amplitude growth rate $\eta$ as a function of frequency $\nu$ of low-degree modes for the model of a $\delta$ Scuti and $\gamma$ Doradus star ($M=1.8\,\mathrm{M}_\odot$, $\log L/\mathrm{L}_\odot=1.3326$, and $\log T_\mathrm{e}=3.8188$). It is clear from Fig.~\ref{fig3} that the pulsation amplitude depends primarily on the frequency of the mode, and has little response with changing $l$. In this work, modes with $l=1$--4 are considered, but the results may also hold for high-degree modes. For solar oscillations in our previous work, similar results were concluded for $l=1$--25 (Xiong \& Deng 2010). The reason is that the distribution of pulsation amplitude, i.e. the eigen-amplitude function, is nearly the same for given frequencies with different $l$.

\subsection{Dependence of pulsation stability on the radial order $n$ of oscillation modes}\label{s34}
\begin{figure}
\includegraphics[width=\columnwidth]{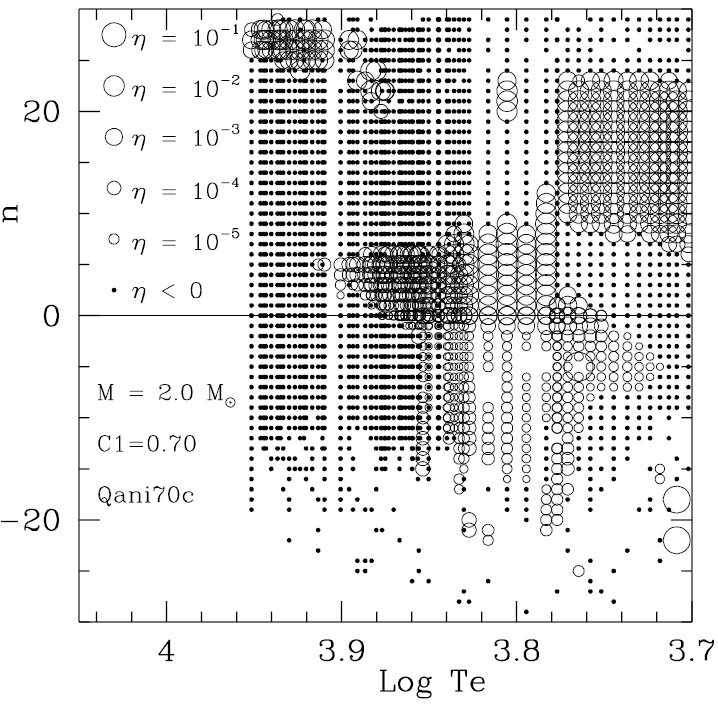}
\caption{Pulsationally stable (small dots) and unstable (open circles) modes in the $n$--$\log T_\mathrm{e}$ plane for the evolutionary models of a $2.0\,\mathrm{M}_\odot$ star, where $n=n_\mathrm{p}-n_\mathrm{g}$ is the radial order of oscillation modes, and $n_\mathrm{p}$ and $n_\mathrm{g}$ are respectively the node number for p and g modes (Unno et al. 1989). The size of the circles is proportional to the amplitude growth rate.}
\label{fig4}
\end{figure}

\begin{figure}
\includegraphics[width=\columnwidth]{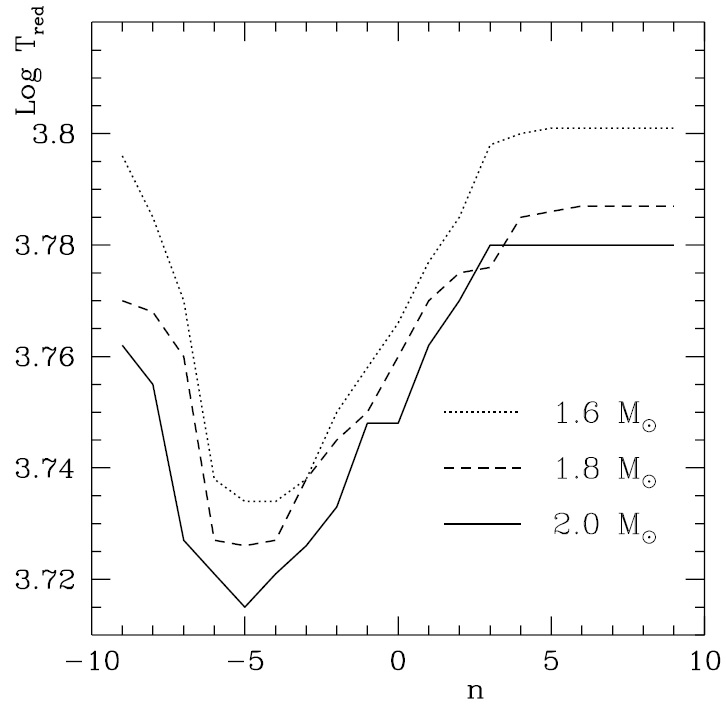}
\caption{The red edge of instability strip versus the radial order of oscillation modes for evolutionary models with $M=1.6\,\mathrm{M}_\odot$ (dotted line), $1.8\,\mathrm{M}_\odot$ (dashed line), and $2.0\,\mathrm{M}_\odot$ (solid line).}
\label{fig5}
\end{figure}

Fig.~\ref{fig4} shows the distribution of pulsationally stable (small dots) and unstable (open circles) modes in the $n$--$\log T_\mathrm{e}$ plane calculated for the evolutionary sequence of a $2.0\,\mathrm{M}_\odot$ star, where $n=n_\mathrm{p}-n_\mathrm{g}$ is the radial order, and $n_\mathrm{p}$ and $n_\mathrm{g}$ are respectively the p-mode and g-mode node number (Unno et al. 1989). $n=0$ is the fundamental mode, $n>$ 0 is for p modes, and $n<$ 0 is for g modes. The size of the circles is proportional to the amplitude growth rate in logarithmic scale. As one can see from Fig.~\ref{fig4}, the blue and red edges of the instability strip are slowly shifting to the blue for higher radial order (larger $n$).  Fig.~\ref{fig5} demonstrates the red edge of unstable region $\log T_\mathrm{r}$ as a function of radial order $n$ for evolutionary models of stars with $M=1.6\,\mathrm{M}_\odot$, $1.8\,\mathrm{M}_\odot$, and $2.0\,\mathrm{M}_\odot$. From Figs.~\ref{fig4} and \ref{fig5}, we can clearly see that the red edge of the instability strip varies with the radial order of modes. It reaches the reddest at $n \sim -5$--$-6$, and then moves towards blue when oscillations shift to higher- or lower-order modes.

\section{Comparison between our theoretical results and other earlier work, and observations}\label{s4}

\begin{figure}
\includegraphics[width=\columnwidth]{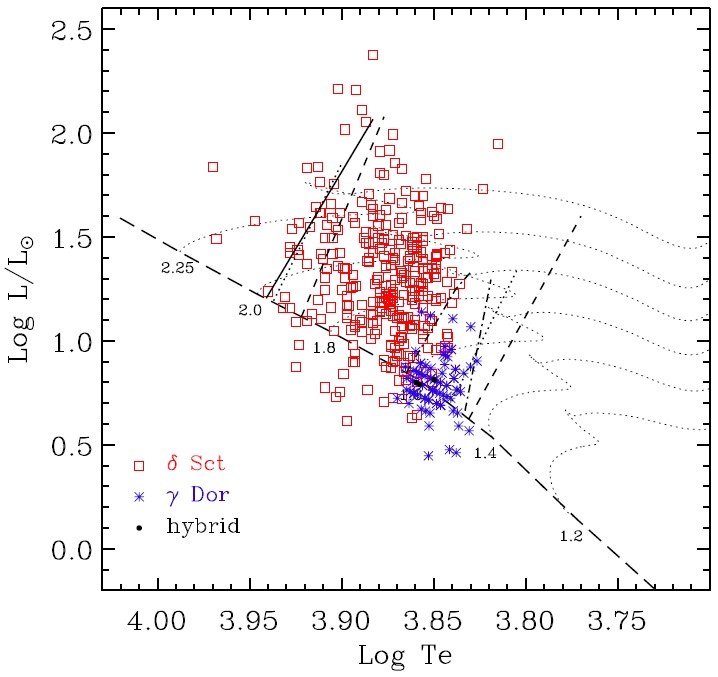}
\caption{H-R diagram of $\delta$ Scuti (open squares), $\gamma$ Doradus (asterisks), and hybrid (filled circles) stars from ground-based observations. The long-dashed line is ZAMS, and the thin dotted lines are evolutionary tracks with masses denoted. The solid line is the theoretical blue edge of the $\delta$ Scuti instability strip derived by Pamyatnykh (2000). The dashed lines represent the instability strip of $\delta$ Scuti stars by Xiong \& Deng (2001). The dotted and dash-dotted lines are respectively the instability strips of $\delta$ Scuti and $\gamma$ Doradus stars by Dupret et al. (2005).}
\label{fig6}
\end{figure}

From ground-based observations, the instability strips of $\delta$ Scuti and $\gamma$ Doradus stars are well defined. In Fig.~\ref{fig6}, the open squares represent a subset of $\delta$ Scuti stars from the catalogue by Rodr\'iguez, L\'opez-Gonz\'alez \& L\'opez de Coca (2000), with parameters derived from Str\"omgren photometry; the asterisks represent confirmed $\gamma$ Doradus field stars compiled by Henry, Fekel \& Henry (2011). The Padova evolutionary tracks (thin dotted lines) and ZAMS (long-dashed line) are also shown. As can be seen, $\delta$ Scuti stars are relatively hot, with masses greater than 1.5\,$\mathrm{M}_\odot$, while $\gamma$ Doradus stars are confined to a small region at lower temperatures near ZAMS. Although the two instability strips overlap, $\delta$ Scuti/$\gamma$ Doradus hybrids from ground-based observations are rare. The filled circles in Fig.~\ref{fig6} are the only three hybrid stars/candidates detected from the ground (Henry et al. 2011).

The theoretical blue edge of the instability strip of $\delta$ Scuti stars has been determined by many authors. The solid line in Fig.~\ref{fig6} shows the result of Pamyatnykh (2000) for radial p1--p8 modes with the mixing-length parameter $\alpha=1.0$. However, the red edge could not be determined using the assumption of frozen-in convection.  By taking into account the interaction between convection and oscillations, the theoretical red edge has been successfully modelled by Houdek (2000), Xiong \& Deng (2001) and Dupret et al. (2005). Although different physical mechanisms were found responsible for stabilizing pulsation, they obtained comparable results that matched observations (Xu et al. 2002; Houdek 2008). In Fig.~\ref{fig6}, the dashed lines mark the theoretical instability strip of $\delta$ Scuti stars derived by Xiong \& Deng (2001) for f--p8 modes, and the dotted lines represent the results of Dupret et al. (2005) for p1--p7 modes with $\alpha=1.8$.

Dupret et al. (2005) also derived the theoretical $\gamma$ Doradus instability strip, which was found to be very sensitive to the adopted $\alpha$ value.  Their results with $\alpha=2.0$ (dash-dotted lines in Fig.~\ref{fig6}) matched observations best. Hybrid stars were predicted in the overlapping region between the $\delta$ Scuti and $\gamma$ Doradus instability strips. In Fig.~\ref{fig2}, our calculations show that the instability strip of $\gamma$ Doradus stars and the overlapping region are much broader. Stars are expected to pulsate simultaneously in p modes and g modes within a large range of $T_\mathrm{e}$.

In the past few years, photometric observations from space missions, especially CoRoT and {\it Kepler}, have greatly improved our understanding of the seismic properties of $\delta$ Scuti and $\gamma$ Doradus stars. The continuous observations with micro-magnitude precision have delivered seismic data of unprecedented quality. Analysis of these data shows the existence of $\delta$ Scuti and $\gamma$ Doradus stars beyond both the blue and red edges of the instability strips (Grigahc\`ene et al. 2010; Uytterhoeven et al. 2011; Hareter 2012; Sarro et al. 2013). Moreover, hybrid stars are common; almost all $\delta$ Scuti and $\gamma$ Doradus stars show hybrid behaviour to some extent (Grigahc\`ene et al. 2010; Balona 2014).

\begin{figure}
\includegraphics[width=\columnwidth]{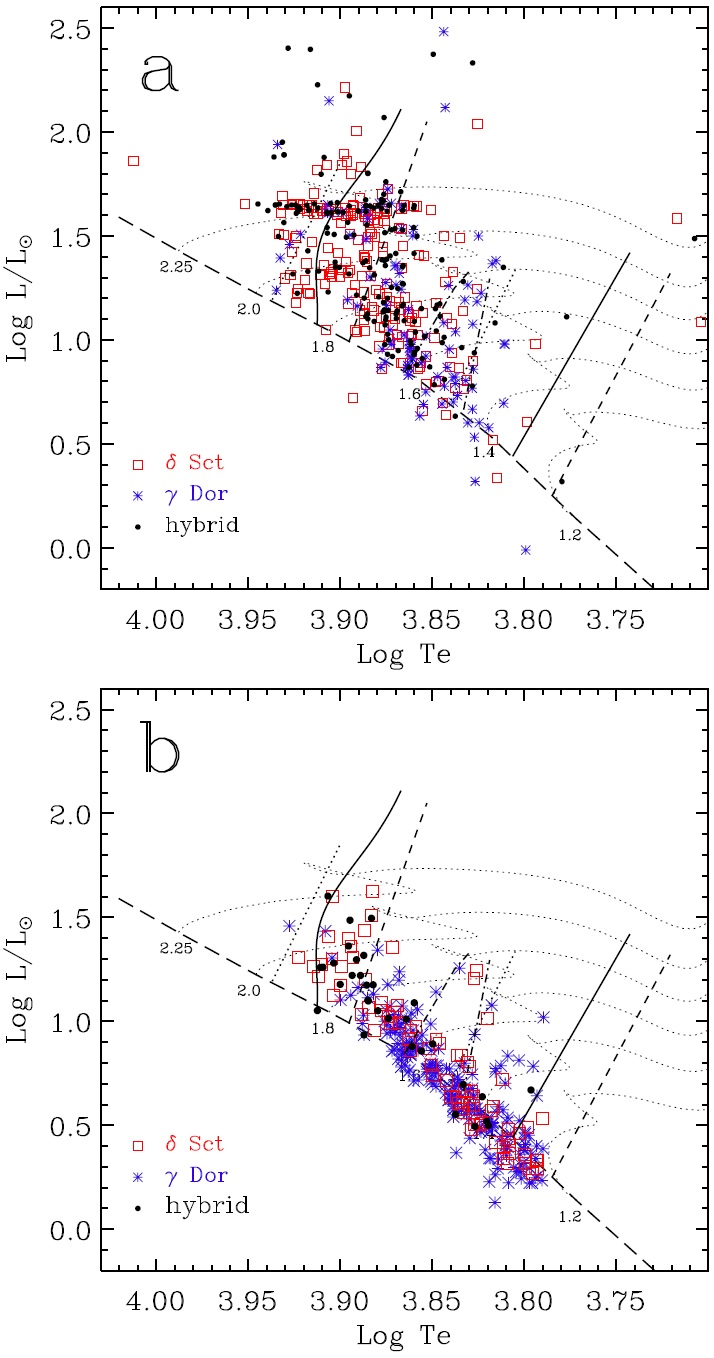}
\caption{H-R diagram of $\delta$ Scuti (open squares), $\gamma$ Doradus (asterisks), and hybrid (filled circles) stars in the {\it Kepler} field identified by Uytterhoeven et al. (2011; panel a) and Bradley et al. (2015; panel b). The long-dashed line is ZAMS, and the thin dotted lines are evolutionary tracks with masses denoted. The solid and dashed lines are respectively the instability strips of $\delta$ Scuti and $\gamma$ Doradus stars in this work. The dotted and dash-dotted lines are respectively the instability strips of $\delta$ Scuti and $\gamma$ Doradus stars by Dupret et al. (2005).}
\label{fig7}
\end{figure}

\begin{figure}
\includegraphics[width=\columnwidth]{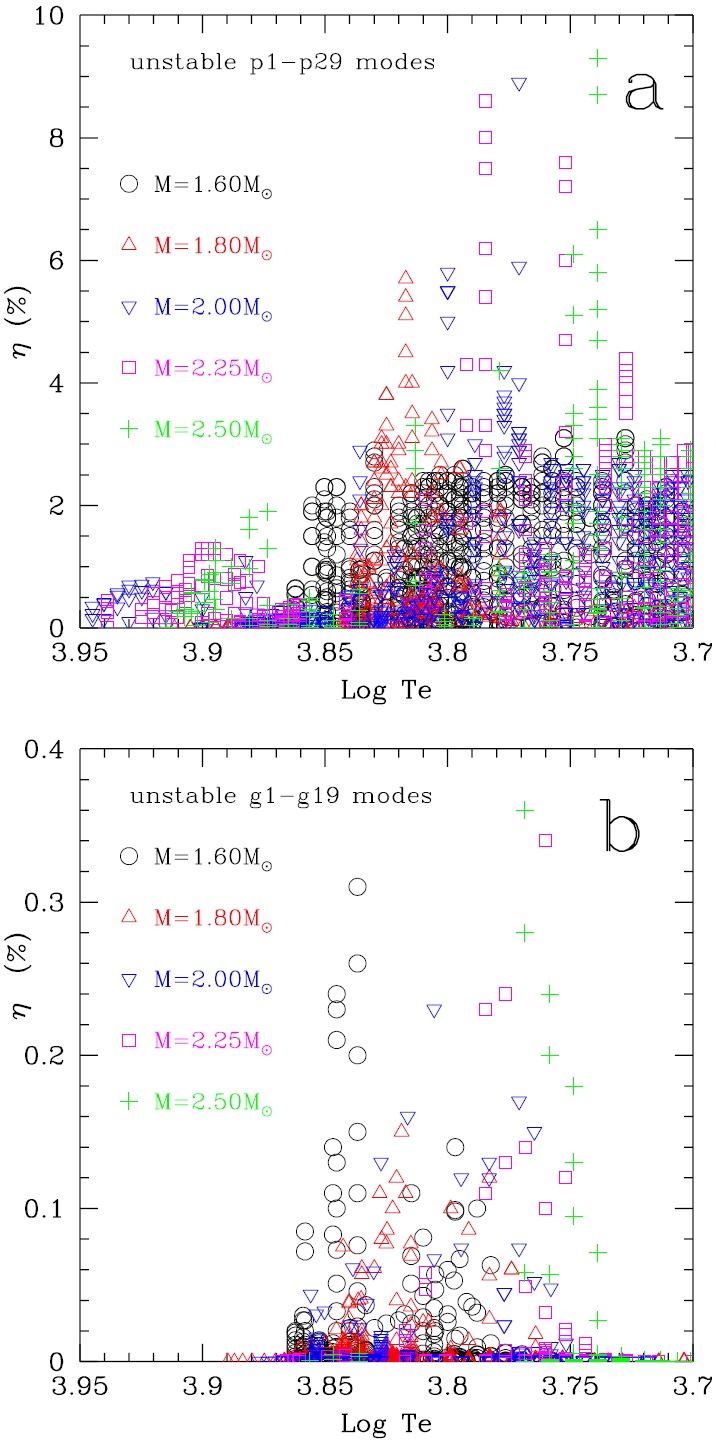}
\caption{Theoretical amplitude growth rate as a function of effective temperature for p modes (panel a) and g modes (panel b) of evolutionary models with $M=1.6\,\mathrm{M}_\odot$ (circles), 1.8\,$\mathrm{M}_\odot$ (triangles), 2.0\,$\mathrm{M}_\odot$ (inverse triangles), 2.25\,$\mathrm{M}_\odot$ (squares), and 2.5\,$\mathrm{M}_\odot$ (pluses).}
\label{fig8}
\end{figure}
\begin{figure}
\includegraphics[width=\columnwidth]{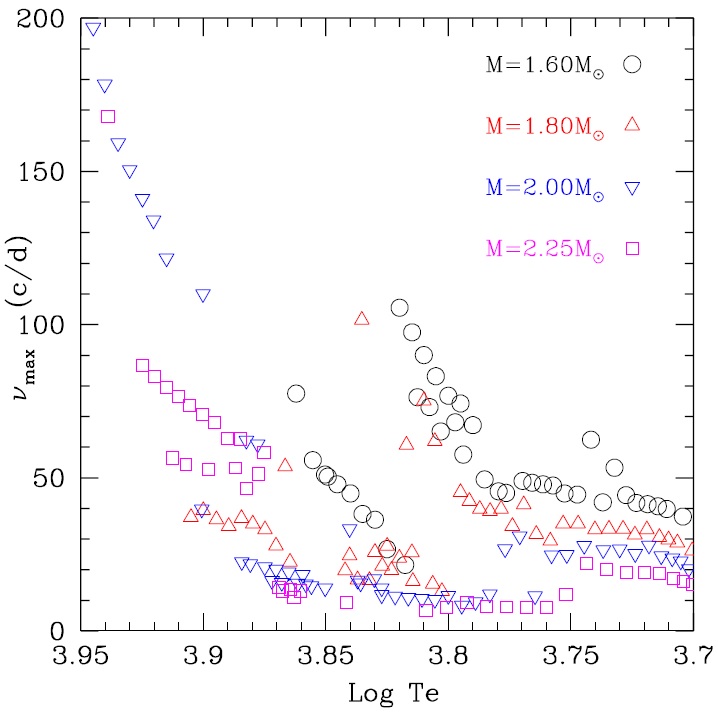}
\caption{Frequency with highest amplitude growth rate of low-degree p modes as a function of effective temperature for theoretical $\delta$ Scuti models with $M=1.6\,\mathrm{M}_\odot$ (circles), $1.8\,\mathrm{M}_\odot$ (triangles), $2.0\,\mathrm{M}_\odot$ (inverse triangles), and $2.25\,\mathrm{M}_\odot$ (squares).}
\label{fig9}
\end{figure}
\begin{figure}
\includegraphics[width=\columnwidth]{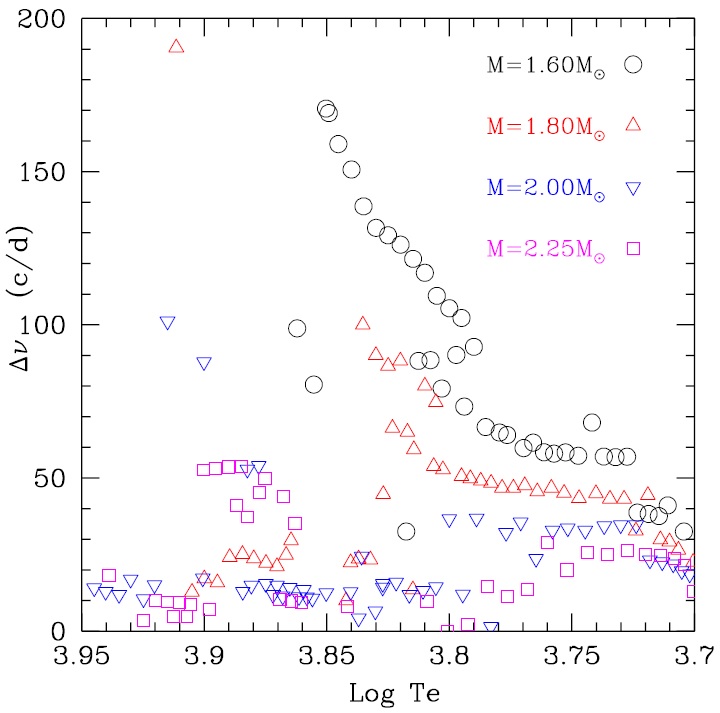}
\caption{Theoretical frequency range of unstable modes for low-degree p modes as a function of effective temperature for stellar models with  $M=1.6\,\mathrm{M}_\odot$ (circles), $1.8\,\mathrm{M}_\odot$ (triangles), $2.0\,\mathrm{M}_\odot$ (inverse triangles), and $2.25\,\mathrm{M}_\odot$ (squares).}
\label{fig10}
\end{figure}

\begin{figure}
\centering
\includegraphics[width=\columnwidth]{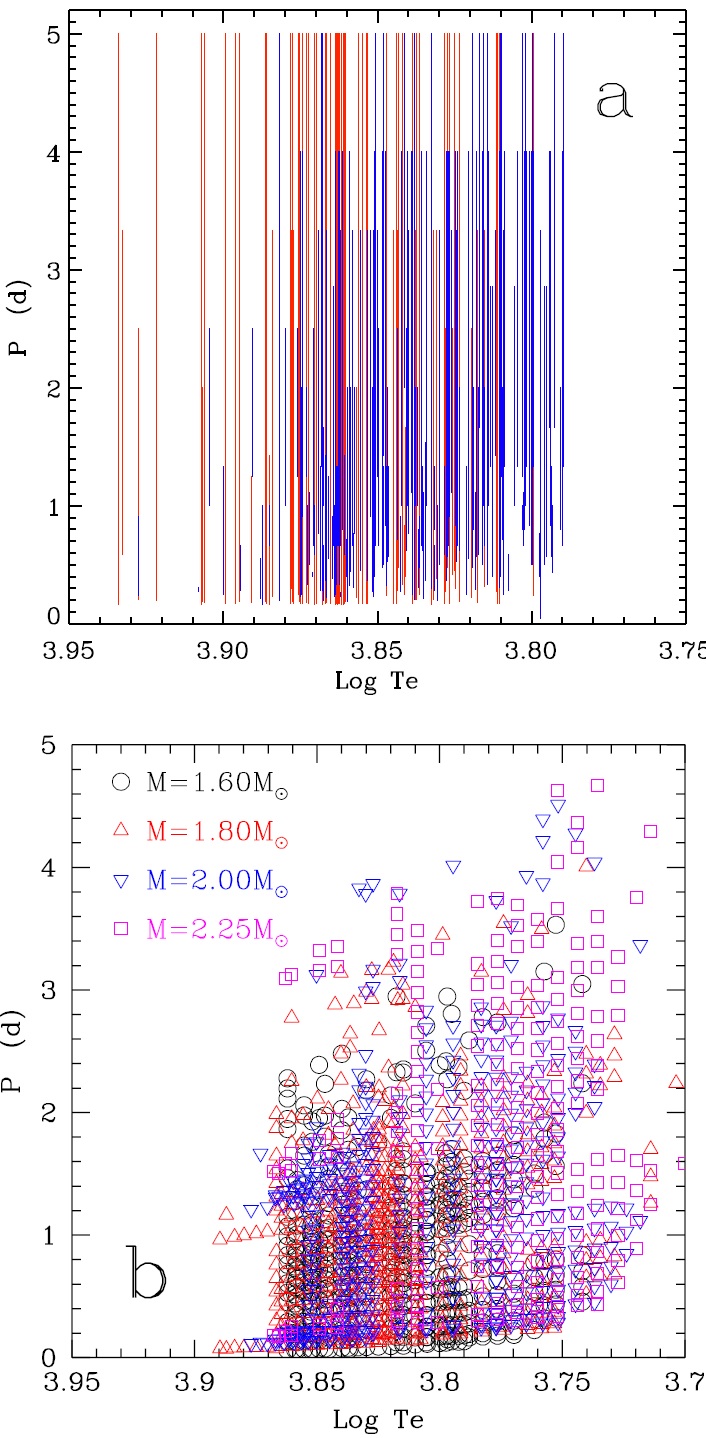}
\caption{Panel a: observed period range of \textit{Kepler} $\gamma$ Dorudas stars as a function of effective temperature. The red lines are $\gamma$ Dorudas stars identified by Uytterhoeven et al. (2011), and the blue lines are $\gamma$ Dorudas stars identified by Bradley et al. (2015). Panel b: theoretical periods of low-degree g modes as a function of effective temperature for stellar models with $M$=$1.6\,\mathrm{M}_\odot$ (circles), $1.8\,\mathrm{M}_\odot$ (triangles), $2.0\,\mathrm{M}_\odot$ (inverse triangles), and 2.25\,$\mathrm{M}_\odot$ (squares).}
\label{fig11}
\end{figure}

\begin{figure}
\centering
\includegraphics[width=\columnwidth]{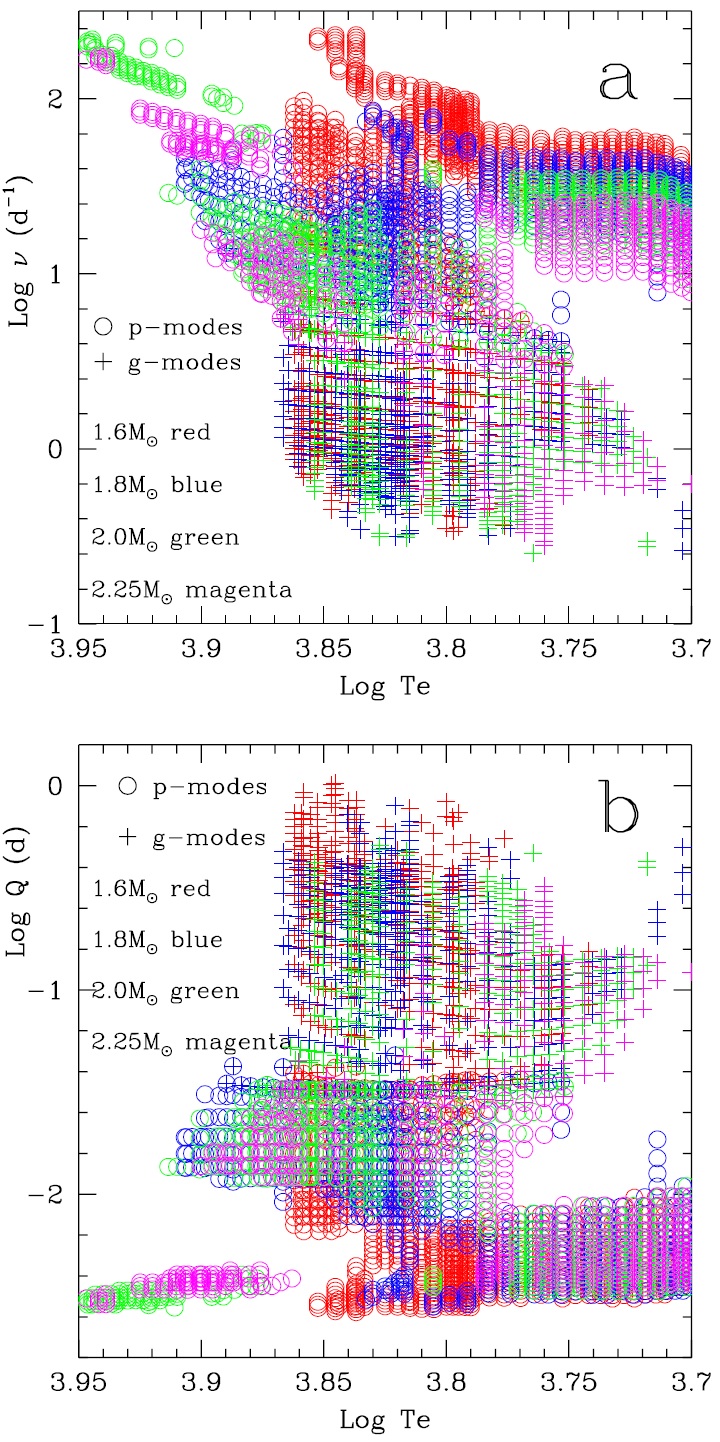}
\caption{Theoretical frequency (panel a) and pulsation constant (panel b) as a function of effective temperature for unstable g (pluses) and p (circles) modes in g20--p29 for models of $M=1.6\,\mathrm{M}_\odot$ (red), $1.8\,\mathrm{M}_\odot$ (blue), $2.0\,\mathrm{M}_\odot$ (green), and 2.25\,$\mathrm{M}_\odot$ (magenta).}
\label{fig12}
\end{figure}

Uytterhoeven et al. (2011) analysed the light curves of all the stars that had been labelled as $\delta$ Scuti or $\gamma$ Doradus candidates in the {\it Kepler} Asteroseismic Science Operations Centre database. They identified 206 $\delta$ Scuti stars, 100 $\gamma$ Doradus stars, and 171 hybrid stars. Fig.~\ref{fig7}a shows the distribution of these stars on the H-R diagram using revised stellar parameters (Huber et al. 2014). It is clear from Fig.~\ref{fig7}a that $\delta$ Scuti and $\gamma$ Doradus stars are no longer confined to previous instability strips as determined from ground-based observations. Especially, $\gamma$ Doradus and hybrid stars occupy a much more extended region, consistent with our results.

There are only a few stars in the low-temperature area in Fig.~\ref{fig7}a. This is mostly caused by the systematic underestimation of $T_\mathrm{e}$ in the {\it Kepler} Input Catalog (Brown et al. 2011), which was used in the target selection. Bradley et al. (2015) analysed 2,768 relatively faint stars in the {\it Kepler} field, and found many cool $\delta$ Scuti, $\gamma$ Doradus, and hybrid stars, as shown in Fig.~\ref{fig7}b. The lower $T_\mathrm{e}$ limit in their target selection was 6,200 K. The existence of $\delta$ Scuti/$\gamma$ Doradus oscillations beyond this limit needs further investigation.

In Fig.~\ref{fig7}, $\gamma$ Doradus stars are in general more concentrated at lower $T_\mathrm{e}$ compared with $\delta$ Scuti stars. There are stars, mostly of $\delta$ Scuti type, that are hotter than the blue edges of the instability strips. Fig.~\ref{fig8}a shows the theoretical amplitude growth rate $\eta$ of p modes as a function of effective temperature for different stellar masses. It is clear that the peak of $\eta$ moves towards lower $T_\mathrm{e}$ as $M$ increases. There are still modes with $\eta>0$ beyond the plotted blue edge, but the amplitude growth rate is generally two orders of magnitude smaller than the peak value. This explains why they have not been observed from the ground. As discussed in Section 3.4, the instability strip moves towards higher temperatures with increasing radial order. The results of Pamyatnykh (2000) and Dupret et al. (2005) showed the same tendency. Therefore these hot stars are most likely pulsating at high radial orders. Balona and Dziembowski (2011) identified 1,568 $\delta$ Scuti stars in the {\it Kepler} field. Their calculations show that many hot $\delta$ Scuti stars in the {\it Kepler} field are pulsating in overtones as high as radial order 8. Their analysis shows that both the observed frequency of maximum amplitude and frequency range increase towards higher $T_\mathrm{e}$ (Figs. 2 and 3 in their paper). Our results of the frequency with highest amplitude growth  rate (Fig.~\ref{fig9}) and frequency range (Fig.~\ref{fig10}) as a function of $T_\mathrm{e}$ show the same trend. Balona and Dziembowski (2011) calculated linear non-adiabatic oscillations for stellar models using standard frozen-in convection. They reached the same conclusion on the frequency with highest growth rate, but failed to predict the observed distribution of frequency range.

Fig.~\ref{fig8}b shows  the theoretical amplitude growth rate of g modes as a function of effective temperature. Like in Fig.~\ref{fig8}a, $\eta$ peaks at $\log T_\mathrm{e}\approx 3.8$. The results of Balona et al. (2011) on $\gamma$ Doradus stars in the {\it Kepler} field showed that the amplitudes of the dominant periodic component have similar distribution but centre at higher temperature (Fig.~7 in their paper).

Fig.~\ref{fig11}a shows the period range of \textit{Kepler} $\gamma$ Doradus stars as a function of effective temperature. The $\gamma$ Doradus stars identified by Uytterhoeven et al. (2011) and Bradley et al. (2015) are shown in red and blue, respectively. Fig.~\ref{fig11}b shows the theoretical periods of unstable g1--g20 modes as a function of effective temperature. From the comparison of Figs.~\ref{fig11}a and b, we can see that the theoretical unstable g modes cover the main period range of \textit{Kepler} $\gamma$ Doradus stars. In this work, the mode search is almost complete for g1--g20 modes (see Fig.~\ref{fig4}).

Fig.~\ref{fig12} shows the theoretical frequency $\nu$ (panel a) and pulsation constant $Q$ (panel b) as a function of effective temperature for unstable g (pluses) and p (circles) modes in g20--p29. Stellar models with $M=1.6\,\mathrm{M}_\odot$, $1.8\,\mathrm{M}_\odot$, $2.0\,\mathrm{M}_\odot$, and $2.25\,\mathrm{M}_\odot$ are shown in red, blue, green, and magenta, respectively. In Fig.~\ref{fig12}a, unstable g modes largely cover the observed frequency range of $\gamma$ Doradus stars, and unstable p modes are in consistent with the observed frequency range of $\delta$ Scuti stars. For stellar models of the same mass, unstable g and p modes are detached in the log\,$\nu-\log T_\mathrm{e}$ diagram. However, for models of different masses, the frequencies of unstable g and p modes partially overlap. In the $\log Q - \log T_\mathrm{e}$ plane (Fig.~\ref{fig12}b), unstable g and p modes become completely separated. For unstable g modes, $Q \gtrsim 0.04\,\mathrm{d}$, corresponding to $\gamma$ Doradus stars, while for unstable p modes, $Q \lesssim 0.04\,\mathrm{d}$, corresponding to $\delta$ Scuti stars.

\section{Excitation and stabilization mechanisms for $\delta$ Scuti and $\gamma$ Doradus stars}\label{s5}

\subsection{Analysis of accumulated work}\label{s51}

\begin{figure}
\includegraphics[width=\columnwidth]{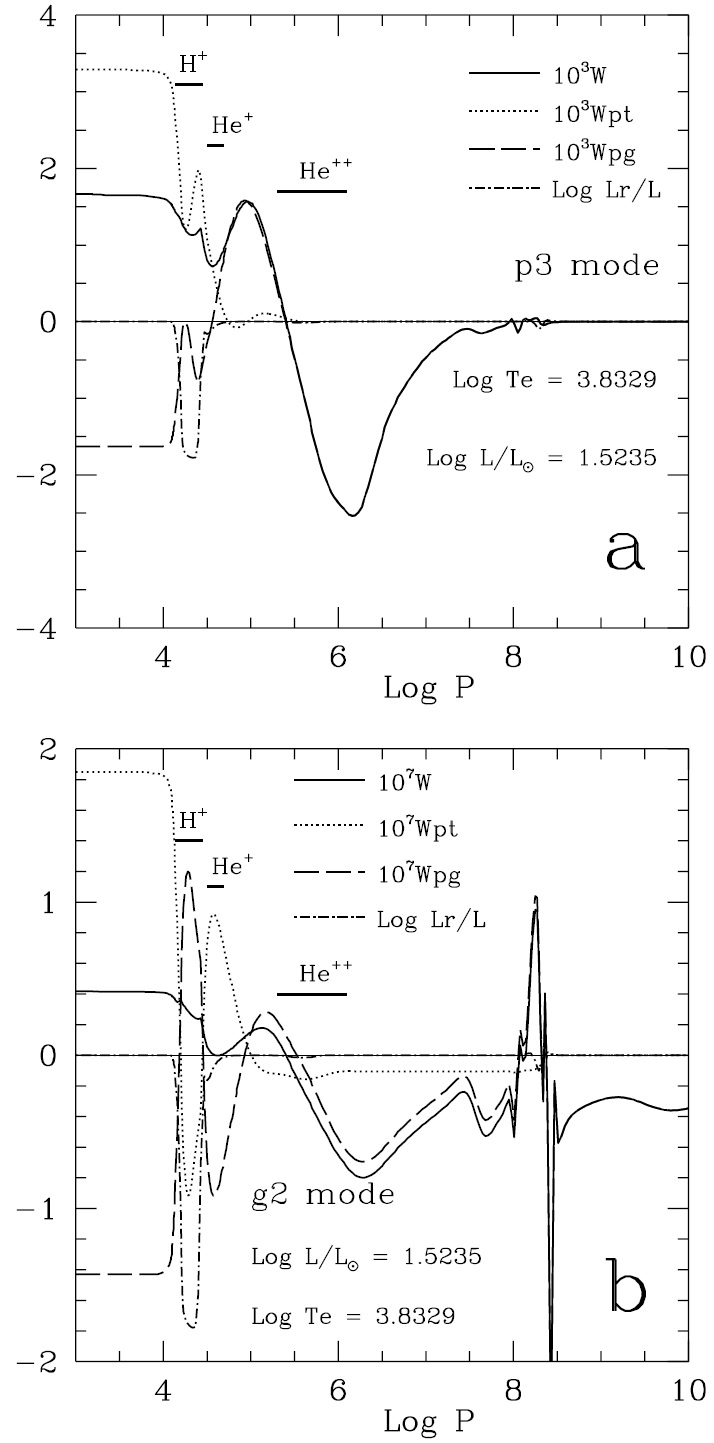}
\caption{Accumulated work versus depth ($\log P$) for a model of warm $\delta$ Scuti/$\gamma$ Doradus hybrid ($M=2.0\,\mathrm{M}_\odot$, $\log L/\mathrm{L}_\odot=1.5235$, $\log T_\mathrm{e}=3.8329$). The solid lines are the total accumulated work $W=W_{P_\mathrm{g}}+W_{P_\mathrm{t}}$. $W_{P_\mathrm{g}}$ (the long-dashed line) and $W_{P_\mathrm{t}}$ (the dotted line) are respectively the gas and turbulent pressure components. The dash-dotted lines are the fractional radiative flux $\log L_r/L$. The horizontal lines indicate the locations of the ionization regions of hydrogen and helium. Panel a: p3 mode. Panel b: g2 mode.}
\label{fig13}
\end{figure}

\begin{figure}
\includegraphics[width=\columnwidth]{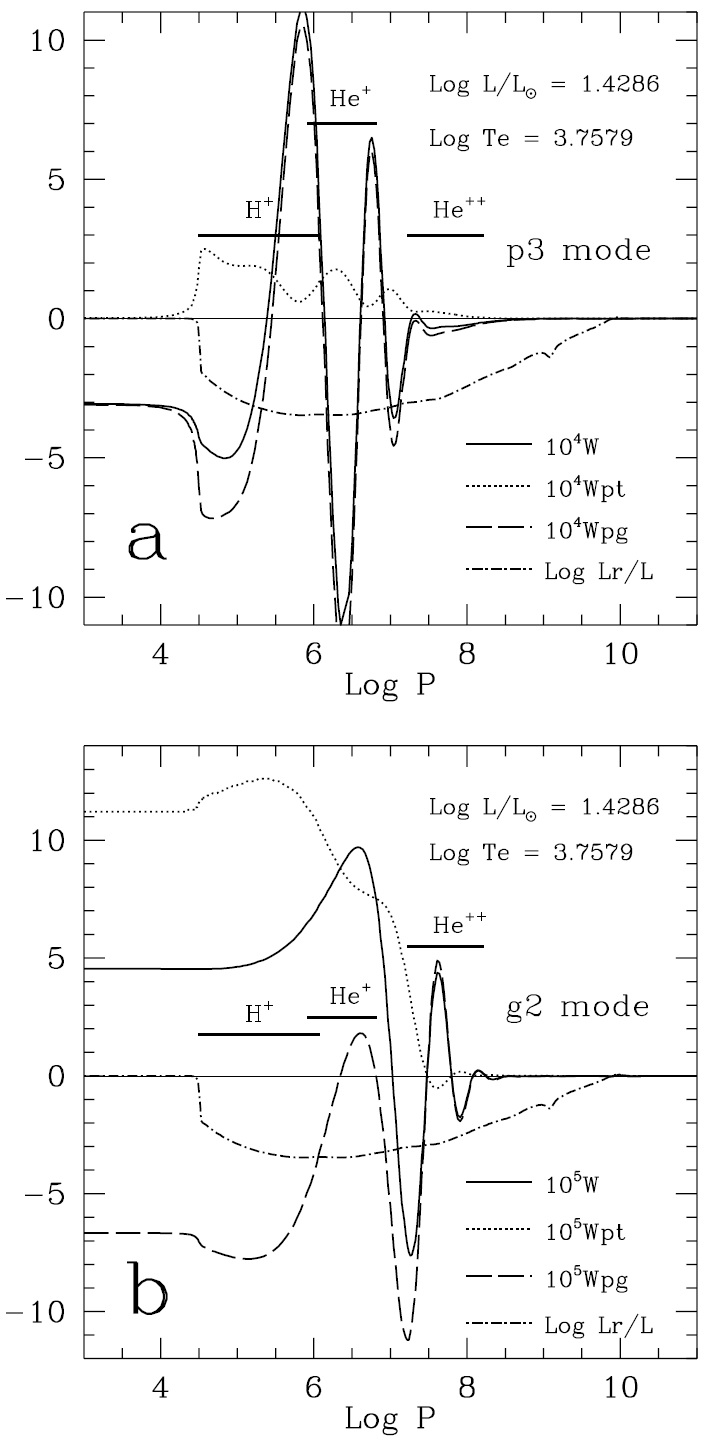}
\caption{Accumulated work versus depth ($\log P$) for a cool star near the red edge of the $\delta$ Scuti-$\gamma$ Doradus instability strip ($M=2.0\,\mathrm{M}_\odot$, $\log L/\mathrm{L}_\odot=1.4286$, $\log T_\mathrm{e}=3.7579$). All the symbols are the same as Fig.~\ref{fig13}. Panel a: p3 mode. Panel b: g2 mode.}
\label{fig14}
\end{figure}

Accumulated work is a simple but efficient way to probe excitation and stabilization mechanisms. Figs.~\ref{fig13}--\ref{fig14} show the accumulated work as a function of depth ($\log P$) in p3 (panel a in each figure) and g2 modes (panel b) respectively in the models of a $2.0\,\mathrm{M}_\odot$ star at two temperatures $\log T_\mathrm{e}=3.8329$ and 3.7579. Where the solid lines are the total accumulated work, $W=W_{P_\mathrm{g}}+W_{P_\mathrm{t}}$, the long-dashed and dotted lines are the gas pressure and turbulent pressure components (Xiong \& Deng 2010), defined as,

\bq
W_{P_\mathrm{g}}=-\frac{\pi}{2E_\mathrm{k}}\int^{M_0}_0{\rm I_m}\left\{\frac{\bar P}{\bar\rho}\frac{\delta\bar P}{\bar P}\frac{\delta\bar\rho^*}{\bar\rho}\right\}dM_r,\label{eq5}
\eq
\begin{align}
W_{P_\mathrm{t}}=&-\frac{\pi}{2E_\mathrm{k}}\int^{M_0}_0{\rm I_m}\left\{2x^2\frac{\delta x}{x}\frac{\delta\bar\rho^*}{\bar\rho}-\chi^{11}\left(\frac{\delta\bar\rho}{\bar\rho}+\frac{\delta\chi^{11}}{\chi^{11}}\right)\right.\nonumber\\
&\left.\times\left[\frac{d}{d\ln r}\left(\frac{\delta r^*}{r}\right)+\frac{l\left(l+1\right)}{2}\frac{\delta r^*_\mathrm{h}}{r}\right]\right\}dM_r,
\label{eq6}
\end{align}
where
\bq
E_\mathrm{k}={1\over2}\int^{M_0}_0\omega^2\left[\delta r^2+l\left( l+1\right)\delta r^2_\mathrm{h}\right]dM_r.\label{eq7}
\eq

The gas pressure component of accumulated work $W_{P_\mathrm{g}}$ contains radiative and convective energy transfer, as well as the contributions due to the conversion between pulsational and turbulent kinetic energy caused by deformation and shear in fluids. Linearization of equation (35) in P1, it is trivial to derive,

\begin{align}
\frac{\delta\bar P}{\bar P}=&\Gamma_1\frac{\delta\bar\rho}{\rho}+\frac{\Gamma_3-1}{\bar P}\bigg\{\bar\rho x^2\bigg(\frac{\delta\bar\rho}{\bar\rho}-3\frac{\delta x}{x}\bigg)\nonumber\\
&-\bar\rho\chi^{11}\bigg[\frac{d}{d\ln r}\bigg(\frac{\delta r}{r}\bigg)+\frac{l( l+1)}{2}\frac{\delta r_\mathrm{h}}{r}\bigg]\nonumber\\
&-\frac{1}{4\pi r^3i\omega}\bigg[\frac{d}{d\ln r}(L'_\mathrm{r}+L'_\mathrm{c}+L'_\mathrm{t})\nonumber\\
&-l(l+1)(L'_\mathrm{r,h}+L'_\mathrm{c,h}+L'_\mathrm{t,h})\bigg ]\bigg\},
\label{eq8}
\end{align}
where $\delta r$ and $\delta r_\mathrm{h}$ are respectively the radial and horizontal components of oscillations, while $\left( L'_\mathrm{r}+L'_\mathrm{c}+L'_\mathrm{t}\right)$ and $\left(L'_\mathrm{r,h}+L'_\mathrm{c,h}+L'_\mathrm{t,h}\right)$ are respectively the radial and horizontal components of radiative flux, convective energy flux and turbulent kinetic energy flux. Inserting equation (\ref{eq8}) into equation (\ref{eq5}), $W_{P_\mathrm{g}}$ can then be rewritten as,

\begin{align}
W_{P_\mathrm{g}}=&\frac{\pi}{2E_\mathrm{k}}\int^{M_0}_0{\rm Im}\bigg\{3(\Gamma_3-1)x^2\frac{\delta x}{x}\frac{\delta\bar\rho^*}{\bar\rho}\nonumber\\
&+(\Gamma_3-1)\chi^{11}\frac{\delta\bar\rho^*}{\bar\rho}\bigg[\frac{d}{d\ln r}\bigg(\frac{\delta r}{r}\bigg)+\frac{l( l+1)}{2}\frac{\delta r_\mathrm{h}}{r_\mathrm{h}}\bigg]\bigg\}dM_r\nonumber\\
&-\frac{\pi}{2E_\mathrm{k}}\int^{M_0}_0{\rm Re}\bigg\{\frac{\Gamma_3-1}{4\pi r^2\bar\rho\omega}\bigg[\frac{d}{d\ln r}(L'_\mathrm{r}+L'_\mathrm{c}+L'_\mathrm{c})\nonumber\\
&-l(l+1)(L'_\mathrm{r,h}+L'_\mathrm{c,h}+L'_\mathrm{t,h})\bigg]\frac{\delta\bar{\rho^*}}{\bar\rho}\bigg\}dM_r.
\label{eq9}
\end{align}

The first integral in equation (\ref{eq9}) represents transformation between oscillation kinetic and thermal energy caused by deformation and shear motion of fluids, which is, when combined with $W_{P_\mathrm{t}}$ of equation (\ref{eq6}), the dynamical coupling between convection and oscillations,

\begin{align}
W_\mathrm{dyn}=&\frac{\pi}{2E_\mathrm{k}}\int^{M_0}_0{\rm Im}\bigg\{(5-3\Gamma_3)x^2\frac{\delta x^*}{x}\frac{\delta\bar\rho}{\rho}+\bigg(\Gamma_3\frac{\delta\bar{\rho^*}}{\bar\rho}\nonumber\\
&+\frac{\delta{\chi^{11}}^*}{\chi^{11}}\bigg)
\bigg[\frac{d}{d\ln r}\bigg(\frac{\delta r}{r}\bigg)+\frac{l(l+1)}{2}\frac{\delta r_\mathrm{h}}{r}\bigg]\bigg\}dM_r.
\label{eq10}
\end{align}

While the second integral in equation (\ref{eq9}) expresses the contributions due to radiative and convective energy transfer,

\begin{align}
W_{L_\mathrm{r}}=&-\frac{\pi}{2E_\mathrm{k}}\int^{M_0}_0{\rm Re}\bigg\{\frac{\Gamma_3-1}{4\pi r^3\bar\rho\omega}\bigg[\frac{dL'_\mathrm{r}}{d\ln r}\nonumber\\
&-l(l+1)L'_\mathrm{r,h}\bigg]\frac{\delta\bar\rho^*}{\bar\rho}\bigg\}dM_r,
\label{eq11}
\end{align}

\begin{align}
W_{L_\mathrm{c,t}}=&-\frac{\pi}{2E_\mathrm{k}}\int^{M_0}_0{\rm Re}\bigg\{\frac{\Gamma_3-1}{4\pi r^3\bar\rho\omega}\bigg[\frac{d}{d\ln r}(L'_\mathrm{c,h}+L'_\mathrm{t,h})\nonumber\\
&-l(l+1)(L'_\mathrm{c,h}+L'_\mathrm{t,h})\bigg]\frac{\delta\bar\rho^*}{\bar\rho}\bigg\}dM_r.
\label{eq12}
\end{align}

The first term in equation (\ref{eq10}) is the isotropic component of Reynold's stress tensor, i.e. the contribution of turbulent pressure. Notice that $5-3\Gamma_3>0$ and ${\rm Im}\left\{\frac{\delta x^*}{x}\frac{\delta\bar\rho}{\bar\rho}\right\}=\left|\frac{\delta x}{x}\right|\left|\frac{\bar\rho}{\bar\rho}\right|\sin\left(\phi_{\delta\rho}-\phi_{\delta x}\right)>0$, therefore the turbulent pressure is always an excitation of oscillations. On the contrary, in the deep interiors of convective zone, convective energy transfer (the thermodynamic coupling between convection and oscillations) is a damping against oscillations in general. This is because of the fact that $L_\mathrm{c}=4\pi r^2\bar\rho\bar{C_P}\bar TV$ (where $V=\overline{w'_rT'}/\bar T$ is the turbulent velocity--temperature correlation), so that

\begin{align}
L'_\mathrm{c}=&L_\mathrm{c}[(A+C_{P,P})\bar P'/\bar P+(1-B+C_{P,T})\bar T'/\bar T\nonumber\\
&+V'/V].
\label{eq13}
\end{align}
Due to the inertia of convective motions, it always lags behind density, therefore $W_{L_\mathrm{c,t}}<0$ normally holds, i.e. the convective energy transfer (or say thermodynamic coupling between convection and oscillations) is a damping mechanism.

$W_{L_\mathrm{r}}$ can be expressed as a summation $W_{L_\mathrm{r}}=W_\mathrm{k}+W_\mathrm{rmd}$, where the two terms are defined as the following:

\begin{align}
W_\mathrm{k}=&-\frac{\pi}{2E_\mathrm{k}}\int^{M_0}_0{\rm Re}\bigg\{\frac{\Gamma_3-1}{4\pi r^3\bar\rho\omega}\bigg[L_\mathrm{r}\frac{d}{d\ln r}\bigg(\frac{L'_\mathrm{r}}{L_\mathrm{r}}\bigg)\nonumber\\
&-l(l+1)L'_\mathrm{r,h}\bigg]\frac{\delta\bar\rho^*}{\bar\rho}\bigg\}dM_r,
\label{eq14}
\end{align}

\bq
W_\mathrm{rmd}=-\frac{\pi}{2E_\mathrm{k}}\int^{M_0}_0{\rm Re}\left\{\frac{\Gamma_3-1}{4\pi r^3\bar\rho\omega}\frac{L'_\mathrm{r}}{L_\mathrm{r}}\frac{dL_\mathrm{r}}{d\ln r}\frac{\delta\bar\rho^*}{\bar\rho}\right\}dM_r.\label{eq15}
\eq
One should notice that,
\begin{align}
\frac{L'_\mathrm{r}}{L_\mathrm{r}}&=-\left(A+K_P\right)\frac{P'}{P}+\left(4+B-K_T\right)\frac{T'}{T}+\frac{d}{d\ln T}\nonumber\\
&\approx -\left[K_T-4-B+\frac{A+K_P}{\nabla_\mathrm{ad}}\right]\frac{T'}{T}.
\label{eq16}
\end{align}
Inserting equation (\ref{eq16}) into equation (\ref{eq15}), the latter can then be expressed approximately as
\begin{align}
W_\mathrm{rmd}\approx &\frac{\pi}{2E_\mathrm{k}}\int^{M_0}_0{\rm Re}\bigg\{\frac{1}{4\pi r^3\bar\rho\omega}\frac{dL_\mathrm{r}}{d\ln r}\bigg[K_T-4-B\nonumber\\
&+\frac{A+K_P}{\nabla_\mathrm{ad}}\bigg]\frac{T'}{T}\frac{\delta T^*}{T}\bigg\}dM_r.
\end{align}

$W_\mathrm{k}$ is exactly the $\kappa$-mechanism of accumulated work, and $W_\mathrm{rmd}$ is what we referred as the radiative modulation excitation (RME) in all our previous work (Xiong et al. 1998). The physics of RME is that, in a region with radiative flux gradient, the flux will be modulated by oscillations and will also oscillate. In a cycle of oscillation, a fraction of radiative energy will be converted into kinetic energy of oscillations, which is why the term RME was named in our studies. It may look like $\kappa$-mechanism as it is also related to opacities, but it has distinct physics which makes it exist only in the radiative region with flux gradient such as the top and bottom of a convective envelope. At the top of convective zone, $K_T-4-B+\left(A+K_P\right)/\nabla_\mathrm{ad}>0$, and $dL_\mathrm{r}/d\ln r>0$,  and so is the product of the terms, therefore the region is an excitation zone of RME. For a star with low enough temperature, the convective zone is very deep (deeper than the second ionization zone of helium). At the bottom of the convection zone, both $K_T-4-B+\left(A+K_P\right)/\nabla_\mathrm{ad}$ and $dL_\mathrm{r}/d\ln r$ are negative; the product of the two is again positive, and the RME behaves also as an excitation.

\begin{figure}
\includegraphics[width=\columnwidth]{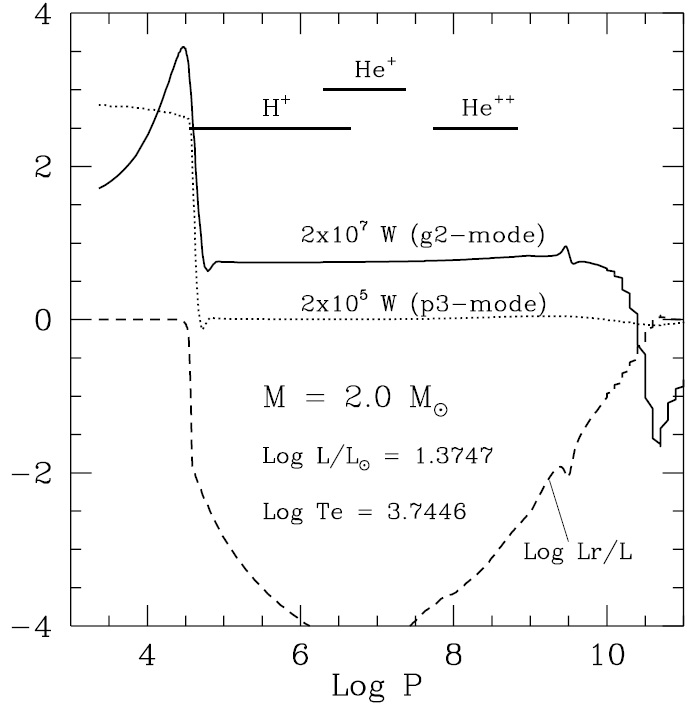}
\caption{Accumulated work versus depth for a cool star outside the $\delta$ Scuti-$\gamma$ Doradus instability strip, but the coupling between convection and oscillations is not taken into account. The solid and dotted lines are respectively g2 and p3 modes.}
\label{fig15}
\end{figure}

RME in our theory has something to do with so-called convective blocking (Guzik et al. 2000), but the difference between them is rather distinct. RME does not need frozen-in convection as assumed in the latter (Warner et al. 2003), neither does it require a convective motion time scale much larger than that of stellar oscillations (Dupret et al. 2004). In any circumstances, the only condition is that $\left[K_T-4-B+\left(A+K_P\right)/\nabla_\mathrm{ad}\right]dL_\mathrm{r}/d\ln r>0$ to have a RME. The other conclusion made for RME is that, when the coupling between convection and oscillations is not taken into account, it is impossible to predict the red edge of the Cepheid and quasi-Cepheid instability strip theoretically. This is due to the fact that, for low-temperature red stars with extended convective envelope, the ionization zones of hydrogen and helium become fully convective, therefore $L_\mathrm{r}/L\ll 1$ and the strength of the $\kappa$-mechanism due to ionization of hydrogen and helium is negligible, $W_k\approx 0$. When neglecting the coupling between convection and oscillations, $W_\mathrm{dyn}=W_{L_\mathrm{c,t}}=0$, as a result $W=W_\mathrm{dyn}+W_{L_\mathrm{c,t}}+W_\mathrm{k}+W_\mathrm{rad}\approx W_\mathrm{rmd}>0$. Fig.~\ref{fig15} depicts the accumulated work for p3 and g2 modes as a function of depth for a star located outside the red edge of the $\delta$ Scuti-$\gamma$ Doradus instability strip. It is shown that, in the radiative zones of radiative flux gradient beyond the upper and lower boundaries of convective region, two prominent bumps emerge, which are due to RME.

After thorough discussions of accumulated work, we turn to work on the excitation and stabilization mechanism of $\delta$ Scuti and $\gamma$ Doradus stars. As the integral of accumulated work is made from the centre outwards, a zone will be an excitation one if accumulated work curve increases toward surface. On the contrary, if the accumulated work decreases outwards, it becomes a damping. Fig.~\ref{fig13} gives the accumulated work of p3 mode (panel a) and g2 mode (panel b) as a function of depth of a $\delta$ Scuti/$\gamma$ Doradus hybrid in the middle of the instability strip. This warm star has only a very shallow hydrogen convective zone; the 1st and 2nd ionization zones of helium are both convectively stable. The layer below the ionized helium region is a radiative damping zone; the primary excitation comes from the $\kappa$-mechanism of the 2nd ionization zone of helium.

Fig.~\ref{fig14} displays the accumulated work of a cool star located near the red edge of the $\delta$ Scuti-$\gamma$ Doradus instability strip. Inside such stars with extended convective envelopes, convection takes over radiation and becomes the main energy transfer mechanism, and the coupling between convection and oscillations then becomes the primary excitation and stabilization mechanism. In previous sections, we have explained that turbulent pressure is in general an excitation of oscillations, while turbulent thermal convection is always a damping against oscillations, and this fact is illustrated clearly in Figs.~\ref{fig13}--\ref{fig14}. When presenting the accumulated work curves in the figures, the upper limits of the integrations in equations (\ref{eq5})--(\ref{eq6}) are regarded as variable $M_r$, therefore the accumulated work represents the total contributions of gas pressure/turbulent pressure $W_{P_\mathrm{g}}\left(M_r\right)$/$W_{P_\mathrm{t}}\left(M_r\right)$ of the whole shell from the bottom boundary $M_r=M_\mathrm{b}$ to $M_r=M_r$. When using stellar surface as the top boundary, the results are $W_{P_\mathrm{g}}\left(M_0\right)$ and $W_{P_\mathrm{t}}\left(M_0\right)$ of the whole stellar envelope $M_\mathrm{b}$--$M_0$, and the surface value (at $W\left(M_0\right)$) of the total accumulated work $W=W_{P_\mathrm{g}}+W_{P_\mathrm{t}}$ equals the amplitude growth rate of oscillations $\eta=-2\pi\omega_i/\omega_r$, where $\omega_i$ and $\omega_r$ are respectively the imaginary and real parts of the complex circular frequency $\omega=\omega_r+i\omega_i$. The calculations of linear non-adiabatic oscillations in this work show that $W(M_0)$ agrees with $\eta$ very well, in most cases the deviation between them is of the order of a few percent. This work justifies in a straightforward way the linear non-adiabatic calculations of oscillations.

As we just explained, no prediction of the red edge of the instability strip of Cepheids and quasi-Cepheid stars can be given if one ignores the coupling between convection and oscillations. The physics is that the radiative $\kappa$-mechanism is solely weakened when neglecting the coupling between convection and oscillations, the RME radiative flux gradient zone in the upper and lower boundaries of a convective region (see Fig.~\ref{fig15}) makes all cool stars always pulsationally unstable. When taking the coupling into account, however, the cool stars become stable, and this then gives rise to the red edge of the instability strip as seen in Fig.~\ref{fig2}. Figs.~\ref{fig14}a and b show the accumulated work for a stellar model with the coupling between convection and oscillations taken into account, which are different from that shown in Fig.~\ref{fig15} when the coupling between convection and oscillations is not considered. The driving around the bottom boundary of the convective zone  has almost disappeared. Such distinct properties are due to the damping effects induced by convective flux (i.e. thermodynamic coupling between convection and oscillations), described by equations (\ref{eq12}) and (\ref{eq15}).

A comparison between the accumulated work of $\delta$ Scuti (panels a) and $\gamma$ Doradus (panels b) stars in Figs.~\ref{fig13}--\ref{fig14} demonstrates that there is hardly any sizable difference between the two types. Therefore it is reasonable to believe that $\delta$ Scuti and $\gamma$ Doradus stars are two subtypes of the same class of variables: $\delta$ Scuti is a p-mode subtype, while $\gamma$ Doradus is a g-mode subtype of oscillators. The corresponding instability strips of the two subtypes actually belong to a greater $\delta$ Scuti--$\gamma$ Doradus instability strip. The $\delta$ Scuti strip is shifted towards blue colours with respect to that of $\gamma$ Doradus within the greater unified instability strip, and the two subtypes largely overlap with each other. In the common region we will find so-called $\delta$ Scuti/$\gamma$ Doradus hybrids. Such a phenomenon resembles the relation between RRc and RRab stars in RR Lyrae variables in which double-mode RRd stars are found in the overlapping area.

There are historical reasons to classify $\gamma$ Doradus stars as a distinct type of oscillators from $\delta$ Scuti stars. For ground-based observations in old times, only the cool $\gamma$ Doradus stars with large amplitudes could be detected. They caught great attentions from observers because of their red colour outside the well-known classical $\delta$ Scuti instability strip and extraordinarily long periods. The long periods inferred that they were oscillating in g modes, instead of typical p modes of $\delta$ Scuti stars. Further analysis based on their location outside the red edge of $\delta$ Scuti instability strip made people believe that they were not excited by traditional $\kappa$ mechanism. This then made convective blocking mechanism (Guzik et al. 2000) a very popular picture. With the groundbreaking discovery of a population of warm $\gamma$ Doradus stars from the {\it Kepler} mission (Balona et al. 2011), convective blocking mechanism should be reviewed, at least it cannot be responsible for the excitation of warm $\gamma$ Doradus stars whose convective envelopes are too shallow for such mechanism to function efficiently, as shown in Fig.~\ref{fig13}b. We are therefore convinced that convective blocking mechanism cannot be the common and only one excitation mechanism for, at least, all $\gamma$ Doradus stars. It may still work only for the classical cool $\gamma$ Doradus stars, if it will ever work as an excitation. Using a time-dependent convection theory, Dupret et al. (2005) calculated oscillations of a $1.6\,\mathrm{M}_\odot$ model, and confirmed that convective blocking mechanism was the driving mechanism in $\gamma$ Doradus. In the current work, we calculated non-adiabatic linear oscillations of stellar models with $M=1.4$--3.0\,$\mathrm{M}_\odot$ using non-local and time-dependent convection theory. We discovered that excitation does not come from the bottom of the convective zones in either warm or cool $\gamma$ Doradus models.  For warm $\gamma$ Doradus stars, the excitation is due to the second ionization zone (Fig.~\ref{fig13}b) instead of convective blocking usually seen as a bump on the accumulated work curve near the bottom of the convective zone.  While for cool ones, there is neither any sign of a bump feature at the bottom of the convective zone (Fig.~\ref{fig14}b), which otherwise is the evidence of convective blocking. All of the numerical results tend to support that there is no substantial difference in excitation mechanism between $\delta$ Scuti and $\gamma$ Doradus, the only difference is that the $\gamma$ Doradus instability strip is slightly offset to a redder colour on the H-R diagram with respect to that of $\delta$ Scuti. The true physical reason responsible is that the oscillations tend to happen in the surface layers for high-order modes. For low-order modes, g modes especially, oscillations have larger amplitudes in deep interior of stars. It is not difficult to understand why the red edge of the instability strip changes with the radial order of oscillation modes as shown in Fig.~\ref{fig5}. The distribution of amplitudes as a function of radial order of oscillation modes leads to the location of instability strip changing also with varying radial order of oscillation modes. Rather, such phenomenon is quite common, such as RR Lyrae stars, where RRc instability strip is observed to be bluer than that of RRab.

\subsection{Radiative damping in deep interior of stars}\label{s52}

Our results on the excitation and classification of $\gamma$ Doradus stars are very different from the commonly accepted interpretations. The main reason leading to such differences is the employment of different convection theories. The coupling between convection and oscillations becomes important in the excitation and stabilization of oscillations in low-temperature $\gamma$ Doradus stars with extended convective envelopes. The theoretical results on the non-adiabatic oscillations depend sensitively on the adopted convection theory. Based on our experience, stellar stability is affected not only by the time-dependent convection in the calculation of non-adiabatic oscillations, but also by the convective non-locality of equilibrium models. Unfortunately, we are unable to carry out evolutionary calculations of intermediate- to low-mass stars under complete non-local convection. Therefore non-local envelope models were used in the calculations of non-adiabatic oscillations instead of full stellar models. This makes little difference in the analysis of p-mode oscillations. However, g-mode oscillations penetrate more deeply into the radiative core, where radiative damping may have non-negligible effects (Dupret et al. 2005). This raises concern about the use of envelope models, which requires further examination.

\begin{figure}
\includegraphics[width=\columnwidth]{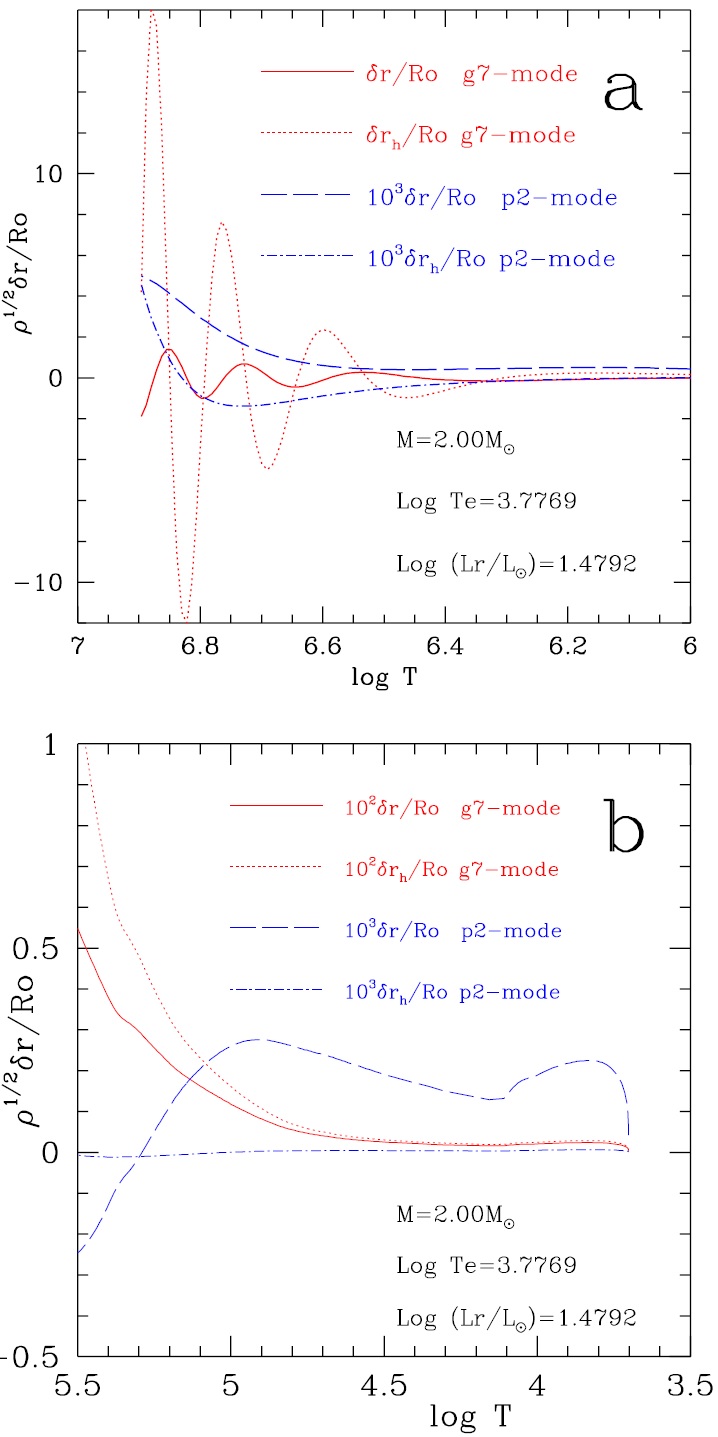}
\caption{Relative amplitude weighted by square root of density as a function of temperature for g7 and p2 modes of a $\delta$ Scuti/$\gamma$ Doradus model ($M=2.0\,\mathrm{M_\odot}$, $\log L/\mathrm{L_\odot}=1.4792$, $\log T_\mathrm{e}=3.7769$) in the core region (panel a) and surface region (panel b). The solid and dotted lines are respectively the radial and horizontal components of the relative amplitude of g7 mode. The dashed and dash-dotted lines are respectively the radial and horizontal components of the relative amplitude of p2 mode.}
\label{fig16}
\end{figure}

Fig.~\ref{fig16} shows the relative amplitude weighted by square root of density as a function of temperature for g7 and p2 modes of a $\delta$ Scuti/$\gamma$ Doradus model. Panel a and b show respectively the core and surface regions, where the relative amplitudes are normalised at the stellar surface ($\delta r/R_0=1$). From the comparison between panel a and panel b, one can see that the relative amplitudes of p2 mode in the core and surface regions are of the same order of magnitude, while the relative amplitude of g7 mode in the core region is 2--3 orders of magnitude higher than in the surface region.

\begin{figure}
\includegraphics[width=\columnwidth]{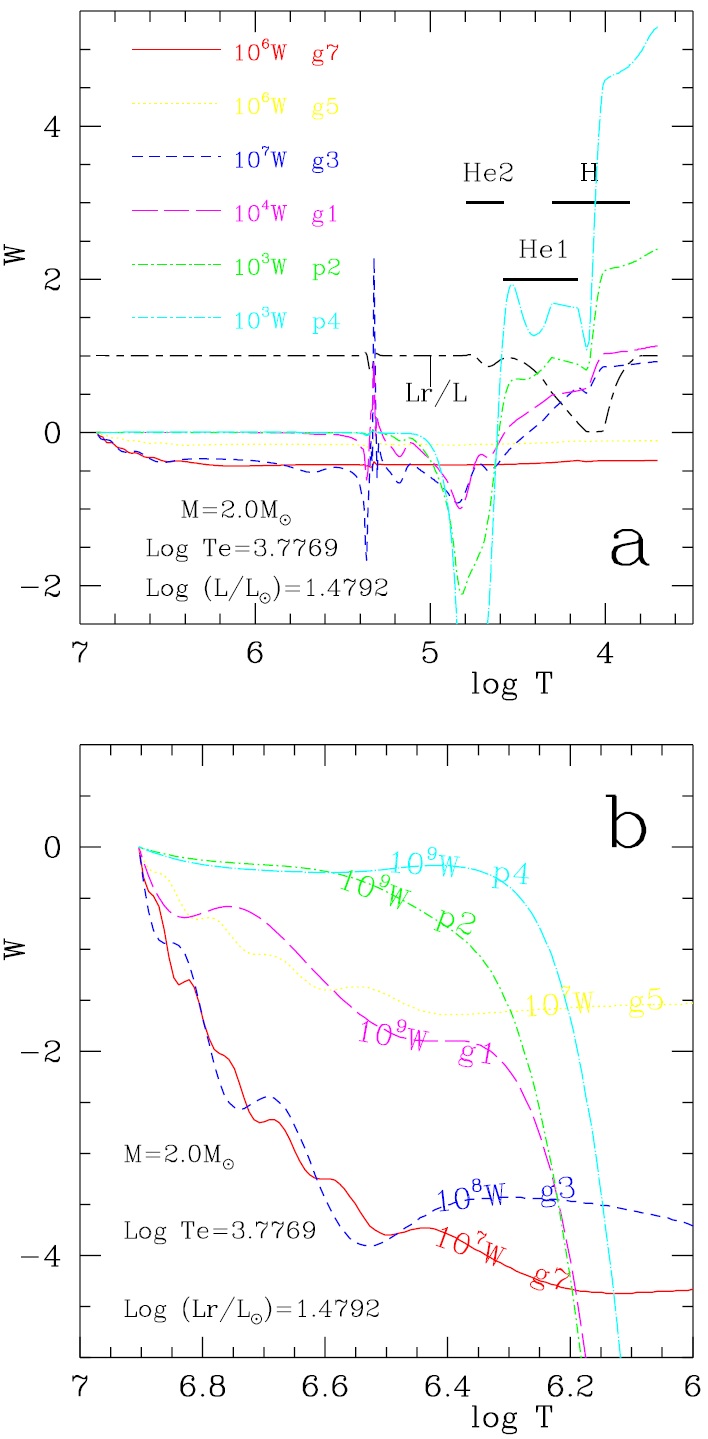}
\caption{Panel a: accumulated work as a function of temperature for g7--p2 modes of the $\delta$ Scuti/$\gamma$ Doradus model as in Fig.~\ref{fig16}. The alternate long and short dashed line is the fractional radiative flux $L_r/L$. The horizontal lines indicate the locations of the ionization regions of hydrogen and helium. Panel b: close-up view of the core region.}
\label{fig17}
\end{figure}

Fig.~\ref{fig17}a shows the accumulated work as a function of temperature for g7--p4 modes of the same stellar model. As can be seen, the pulsationally stable high-order (g3--g7) modes all have small amplitude growth rates; the damping mechanism is radiative damping from the deep interior. However, g1 and p2 modes are unstable, and their amplitude growth rates are 2--3 orders of magnitude higher than those of g3--g7 modes, showing no visible radiative damping in the core in Fig.~\ref{fig17}a. A close-up inspection of the core region (Fig.~\ref{fig17}b) shows that there is radiative damping in the accumulated work of g1 and p2 modes, which is of the same order of magnitude ($\sim 10^{-8}$) compared with high-order g modes. It is similar to the result of Dupret et al. (2005) using full stellar models. As can be seen in Fig.~\ref{fig17}a, the excitation of all unstable g and p modes comes from the stellar surface region. The accumulated work is several orders of magnitude higher than radiative damping in the deep core region. Radiative damping has almost no substantial influence on the pulsational stability of these unstable modes.

It is worth pointing out that there is a sharp variation in the accumulated work at $\log T \sim 5.4$ in Fig.~\ref{fig17}a, resulting from the existence of a narrow convection zone at the iron opacity bump (see also the small dip in $L_r/L$). It causes perturbations in numerical calculations. The perturbation in the accumulated work becomes clear when the amplitude growth rate $|\eta|$ of the oscillation mode is small. $|\eta|$ is normally large for most p modes, thus the perturbation is hardly visible, as in p2 mode in Fig.~\ref{fig17}a. The perturbation has little effect on the results of numerical calculations of non-adiabatic oscillations, because the positive and negative parts almost cancel out, leaving nearly zero net work.

\subsection{Effect of bottom depth on pulsation stability}\label{s53}

\begin{figure}
\includegraphics[width=\columnwidth]{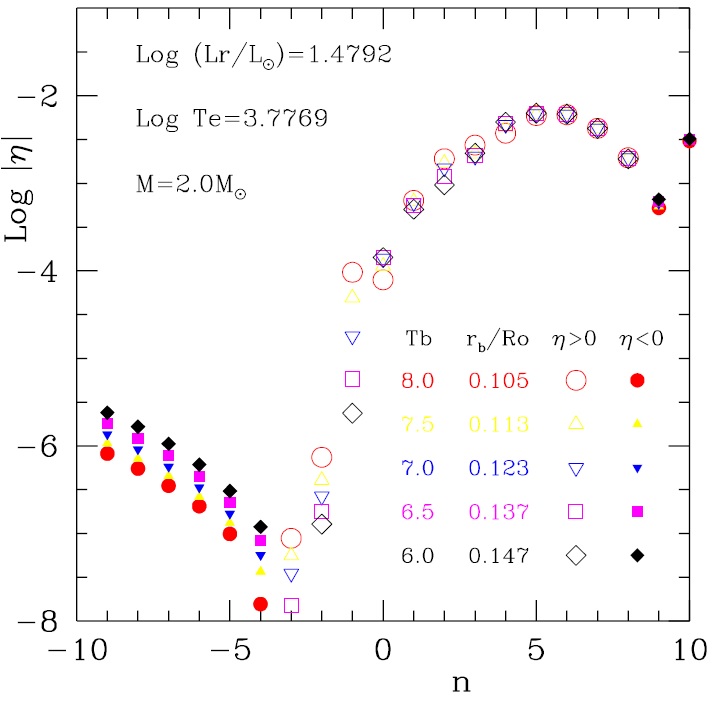}
\caption{Theoretical amplitude growth rate in logarithmic scale $\log |\eta|$ versus the radial order $n$ for the $2.0\,\mathrm{M_\odot}$ stellar model with varying bottom depth. ($T_\mathrm{b}$, $r_\mathrm{b}/R_0$)=($8.0\times 10^6$\,K, 0.105) (circles), ($7.5\times 10^6$\,K, 0.113) (triangles), ($7.0\times 10^6$\,K, 0.123) (inverse triangles), ($6.5\times 10^6$\,K, 0.137) (squares), and ($6.0\times 10^6$\,K, 0.147) (diamonds). Open symbols are unstable modes ($\eta>0$), and filled symbols are stable modes ($\eta<0$).}
\label{fig18}
\end{figure}

In order to evaluate the effects of the depth of the bottom boundary on the amplitude growth rate, we calculated linear non-adiabatic oscillations of the same $2.0\,\mathrm{M_\odot}$ non-local envelope model with bottom boundary set at different depth $T_\mathrm{b}=8.0$--$6.0 \times 10^6$\,K. Fig.~\ref{fig18} shows the amplitude growth rates of g9--p10 modes in logarithmic scale as a function of the radial order, where unstable modes are plotted using open symbols, and stable modes using filled symbols. From Fig.~\ref{fig18} we can draw some clear conclusions.

\begin{figure}
\includegraphics[width=\columnwidth]{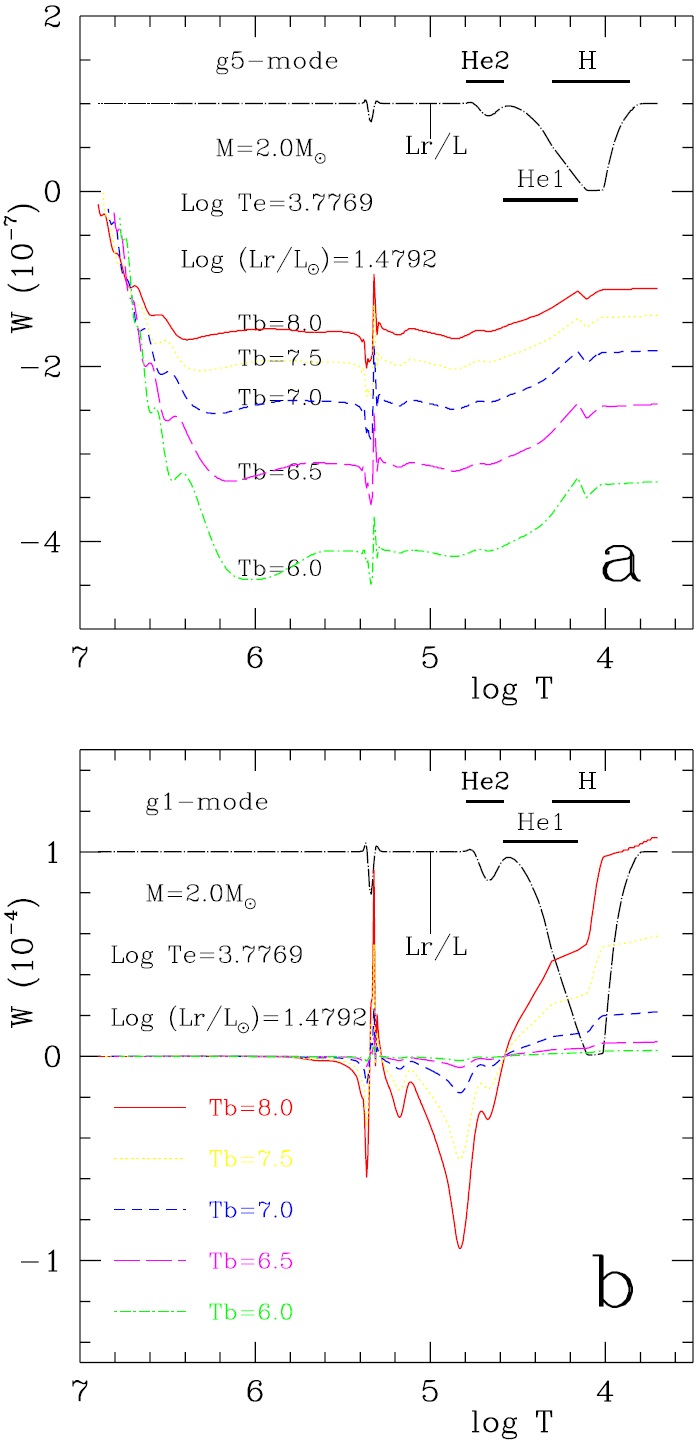}
\caption{Accumulated work as a function of temperature for the $2.0\,\mathrm{M_\odot}$ stellar model with varying bottom depth $T_\mathrm{b}=8.0\times 10^6$\,K (solid line), $7.5\times 10^6$\,K (dotted line), $7.0\times 10^6$\,K (dashed line), $6.5\times 10^6$\,K (long-dashed line), and $6.0\times 10^6$\,K (dash-dotted line). The fractional radiative flux $Lr/L$ and locations of ionization regions are also shown. Panel a: g5 mode. Panel b: g1 mode.}
\label{fig19}
\end{figure}

\begin{enumerate}
\item The amplitude growth rates of p modes are basically unaffected by the change of bottom boundary depth, because the excitation and damping of p modes come mainly from stellar outer layers. Changing the bottom boundary has little influence on their eigenfunctions.
\item The amplitude growth rates of g modes are more sensitive to the change of the bottom boundary depth. As the bottom boundary moves outward, the amplitude growth rates of unstable g modes decrease, and the damping of stable g modes increases. This is because the excitation of unstable g modes still comes from stellar outer layers, and far exceeds radiative damping from the deep interior, while stable g modes are dominated by radiative damping from the deep region. When radiative damping in the interior overtakes the excitation from the surface, the star becomes stable against this mode (Fig.~\ref{fig19}a). When the bottom boundary moves outward, the whole resonance cavity moves outward, and the eigenfunction is also displaced outward. As a result, the damping in the deep region increases (Fig.~\ref{fig19}a), while the excitation from the outer layers decreases (Fig.~\ref{fig19}b).
\item Although the amplitude growth rate changes in value as the bottom boundary moves, the stability or instability of a mode does not change.
\end{enumerate}

\begin{figure}
\includegraphics[width=\columnwidth]{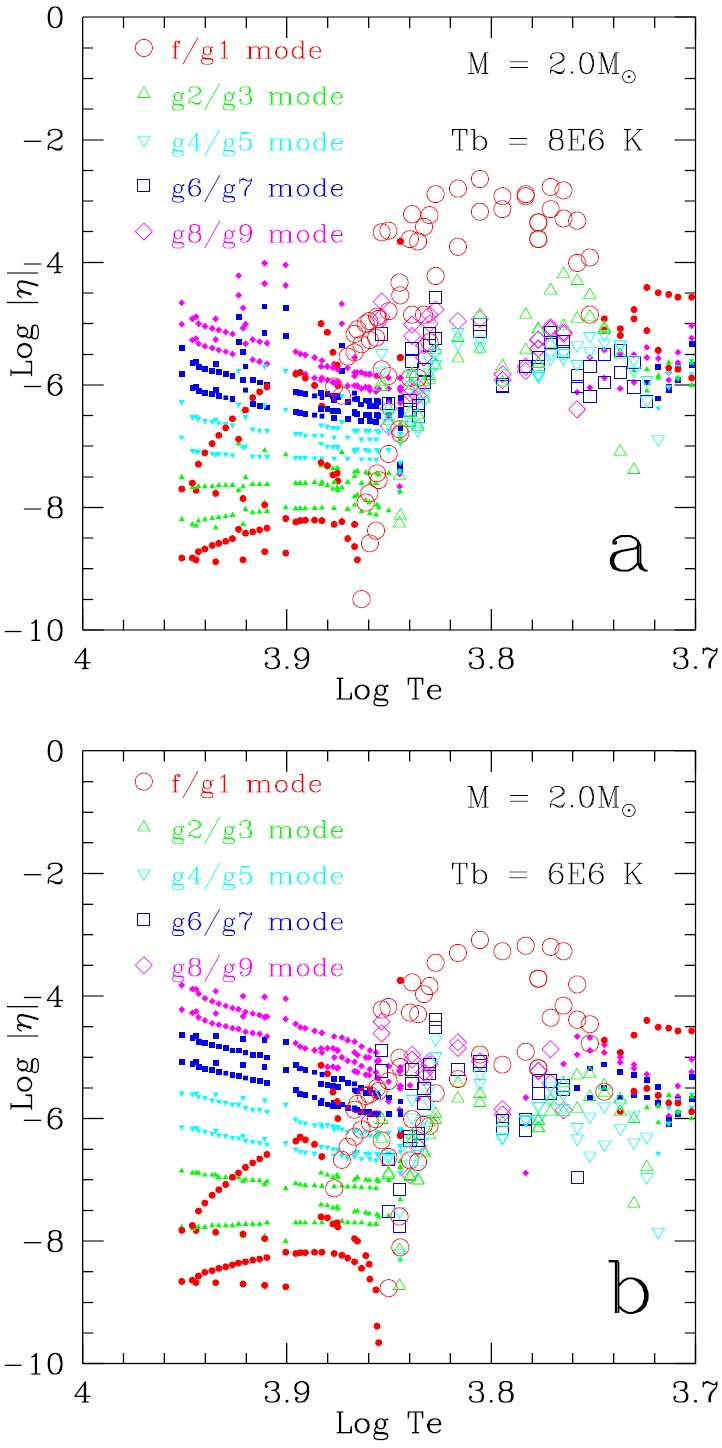}
\caption{Theoretical amplitude growth rates of f--g9 modes in logarithmic scale as a function of effective temperature for an evolutionary sequence of $2.0\,\mathrm{M_\odot}$ models. Open symbols are unstable modes ($\eta>0$), and filled symbols are stable modes ($\eta<0$). Panel a: the bottom boundary is at $T_\mathrm{b}=8\times 10^6$\,K. Panel b: the bottom boundary is at $T_\mathrm{b}=6\times 10^6$\,K.}
\label{fig20}
\end{figure}

Although based on the analysis of a specific $\gamma$ Doradus/$\delta$ Scuti model, these conclusions
generally hold. Fig.~\ref{fig20} shows the amplitude growth rates of f--g9 modes as a function of effective temperature for a sequence of $2.0\,\mathrm{M_\odot}$ envelope models along the evolutionary track with bottom boundary set at $T_\mathrm{b}=8\times 10^6$\,K (panel a) and $T_\mathrm{b}=6\times 10^6$\,K (panel b). The comparison of panel a and panel b shows that all the models follow these same rules.

\section{Summary and discussions}\label{s6}
Using a non-local and time-dependent convection theory, we have calculated radial and low-degree non-radial oscillations of the convective envelopes of stellar evolutionary models with $M=1.4$--$3.0\,\mathrm{M}_\odot$. The main results can be summarised as follows.

\begin{enumerate}

\item The linear stability analysis of g20--p29 modes is almost complete for stellar models with $M=1.4$--$3.0\mathrm{M}_\odot$. In the $\log Q - \log T_\mathrm{e}$ plane, unstable g and p modes form two detached areas. The boundary between them is at $Q \approx 0.04$. For all unstable g modes, $Q \gtrsim 0.04$, while for all unstable p modes, $Q \lesssim 0.04$.

\item On the H-R diagram, unstable g and p modes are located in two partially overlapped instability strips. The g-mode instability strip is systematically redder than the p-mode instability strip.

\item From the analysis of accumulated work, we find no distinct difference in the excitation and stabilization between p and g modes. They are both due to the combination of $\kappa$ mechanism and the coupling between convection and oscillations. Radiative $\kappa$ mechanism plays a major role in the excitation of warm $\delta$ Scuti and $\gamma$ Doradus stars, while the coupling between convection and oscillations is responsible for excitation and stabilization in cool stars.

\end{enumerate}

Based on aforementioned results of linear non-adiabatic oscillations, it seems reasonable to believe that there is no essential difference between $\delta$ Scuti and $\gamma$ Doradus stars. They are just two subgroups of one broader type of $\delta$ Scuti-$\gamma$ Doradus stars below the Cepheid instability strip: $\delta$ Scuti is a p-mode subgroup, while $\gamma$ Doradus is a g-mode subgroup. Within the instability strip, most of the variable stars may be hybrids pulsating in both p and g modes. The connections among them are very similar to that among RR$_\mathrm{ab}$, RR$_\mathrm{c}$, and RR$_\mathrm{d}$.

Clearly, our interpretation of the excitation and stabilization is quite different from the mainstream view as a result of the employment of the non-local time-dependent convection theory. Moreover, we used non-local convection envelope models in the calculation of non-adiabatic oscillations, instead of full stellar models. This is because the effects of non-local convection is crucial to the pulsational stability of stars. These effects exist not only in the time-dependent convection in the calculation of oscillations, but also in the calculation of equilibrium structure of convective envelopes. We carefully studied the influences of radiative damping in stellar deep interior and the depth of bottom boundary on the stability calculations of stellar oscillations. Our results show that:

\begin{enumerate}

\item The excitation and damping of all p modes and unstable g modes come mainly from the outer layers of stars. Radiative damping in the deep region is relatively small, therefore its influence on the pulsational stability is negligible.
\item Radiative damping in stellar deep regions is the main damping mechanism of stable g modes.
\item As the bottom boundary moves outward, the amplitude growth rates of unstable modes decrease, and the damping of stable modes increases. However, the pulsational stability or instability of a specific mode does not change. If the bottom boundary is not set deep enough, the excitation from the outer layers will be underestimated in the calculations of non-adiabatic oscillations, while the radiative damping in the deep region will be overestimated. As a result, the amplitude growth rates decrease, but do not change signs.

\end{enumerate}

All the results in this work are based on chemically uniform envelope models. Unless there exists unusually large damping in the nuclear-reacting core, it is not likely that the use of full stellar models will cause radial changes to the results. We expect to carry out further examinations with full models in near future.

\section*{Acknowledgements}
We are grateful to the referee M.-A. Dupret for helpful comments and suggestions. This work is supported by National Natural Science Foundation of China (NSFC) through grants 11373069, 11473037, and 11403039. CZ acknowledges support from the Young Researcher Grant of National Astronomical Observatories, Chinese Academy of Sciences.







\bsp	
\label{lastpage}
\end{document}